\documentclass[preprint,nofootinbib]{revtex4-1}
\usepackage{graphicx}
\usepackage{dcolumn}
\usepackage{bm}
\usepackage{amsmath}
\usepackage{latexsym}
\usepackage{bbm}
\newcommand{\non}{\nonumber}
\newcommand{\del}{\partial}

\newcommand{\Vac}[1]{\bigg\langle{#1}\bigg\rangle}

\newcommand{\rpartialupx}{\stackrel{\rightarrow}{\raisebox{-0.0ex}{$\partial^\mu_x$}}}

\newcommand{\lpartialupy}{\stackrel{\leftarrow}{\raisebox{-0.0ex}{$\partial^\mu_y$}}}

\def\gtsim{\mathrel{\hbox{\raise0.2ex
\hbox{$>$}\kern-0.75em\raise-0.9ex\hbox{$\sim$}}}}
\def\ltsim{\mathrel{\hbox{\raise0.2ex
\hbox{$<$}\kern-0.75em\raise-0.9ex\hbox{$\sim$}}}}

\begin{document}


\title{$Z'$-ino-driven Electroweak Baryogenesis in the UMSSM}

\author{Eibun Senaha}%
 \email{senaha@eken.phys.nagoya-u.ac.jp}
\affiliation{Department of Physics, Nagoya University, Nagoya 464-8602, Japan}
\bigskip

\date{\today}

\begin{abstract}
The viable window to electroweak baryogenesis in a supersymmetric $U(1)'$ model is studied 
in light of the 126 GeV Higgs boson.
To investigate the decoupling of the sphaleron process in the broken phase, 
we evaluate the sphaleron rate and order of the electroweak phase transition.
In this model, the electroweak phase transition is strongly first order due to the doublet-singlet Higgs mixing. Consequently, for typical parameter sets
the $Z'$ boson has to be lighter than $(150$-$300)$ GeV 
and thus leptophobic to be consistent with the collider bounds.
We also estimate the baryon asymmetry of the Universe based on the closed-time-path formalism,
and find that the $CP$-violating source term fueled by the $Z'$-ino can generate sufficient 
baryon asymmetry.

\end{abstract}

\pacs{Valid PACS appear here}

\maketitle

\section{Introduction}
From cosmological observations such as the cosmic microwave background (CMB),
the big-bang-nucleosynthesis (BBN) etc, the baryon-to-photon ratio in our universe
is found to be~\cite{Beringer:1900zz}
\begin{align}
\eta_{\rm CMB}  &= \frac{n_B}{n_\gamma} = (6.23\pm0.17)\times 10^{-10}, \\
\eta_{\rm BBN}  &= \frac{n_B}{n_\gamma} = (5.1-6.5)\times 10^{-10},~(95\%~{\rm C.L.}).
\end{align}
The two values nicely agree with each other although their relevant time scales
are different.
Revealing the origin of the baryon asymmetry of the universe (BAU) 
is one of the grand puzzles in particle physics and cosmology.
It is well known that the BAU can arise dynamically if the so-called Sakharov's conditions are fulfilled~\cite{Sakharov:1967dj}:
(1) baryon number ($B$) violation, (2) $C$ and $CP$ violation, and (3) departure from thermal equilibrium.
To be consistent with the observations, the BAU must arise after inflation if it exists 
and before the BBN era.

Sakharov's conditions can be satisfied in many ways.
Among others, electroweak baryogenesis (EWBG)~\cite{ewbg} is an attractive framework
since it is intimately related to physics that can be probed by experiments within our reach.
The baryogenesis scenario in the standard model (SM) is based on such an EWBG mechanism.
However, it turns out that the magnitude of the $CP$-violating effect 
coming from the Cabibbo-Kobayashi-Maskawa matrix is far too small~\cite{ewbg_sm_cp}, 
and the electroweak phase transition (EWPT) is a smooth crossover~\cite{sm_ewpt} so that
the EWBG mechanism does not work in the SM.
This shortcoming of the SM motivates one to search for new physics.
The EWBG in the minimal supersymmetric SM (MSSM) has been well studied so far.
However, a light stop scenario, in which the EWPT can be 
strongly first-order, is currently in tension with the Large Hadron Collider (LHC) data~\cite{MSSM-EWBG_LHCtension}.
This difficulty may be circumvented in the extended MSSM models such as 
next-to-MSSM (NMSSM)~\cite{EWBG_NMSSM,Carena:2011jy,Funakubo:2005pu}, 
nearly MSSM (nMSSM)~\cite{Menon:2004wv,Huber:2006wf}, 
$U(1)'$-extended MSSM (UMSSM)~\cite{EWBG_UMSSM}
and its secluded version (sMSSM)~\cite{EWBG_sMSSM,Chiang:2009fs} since the light stop
is not necessarily required to have the strong first-order EWPT.
In the NMSSM, the so-called type-B EWPT~\cite{Funakubo:2005pu} corresponds to such a case.
To realize it, $\lambda\gg\kappa$ may be needed, where $\lambda$ denotes
the coupling constant between the Higgs doublet and singlet fields, and $\kappa$ is 
a trilinear coupling constant of the singlet Higgs field.
In Ref.~\cite{Cheung:2012pg}, it is demonstrated that the superpartner of the singlet Higgs boson, 
which is called the singlino,
can have a significant $CP$-violating effect that can generate the sufficient BAU,
which is called singlino-driven EWBG.
In this case, to avoid sizable baryon washout by the strong sphaleron, the lighter stop and/or sbottom 
should be lighter than about 500 GeV. 
The reason is that the light colored particles can modify a numerical factor, $r_1$ (defined in Eq.~(\ref{r1r2})),
so the the washout by strong sphaleron can be inefficient.
However, the chosen parameter space is in tension with the stop and sbottom searches at the LHC.
If the Higgsino and singlino masses were degenerate, a resonance enhancement
in the $CP$-violating source term would compensate the strong sphaleron washout 
and thus the LHC constraints could be evaded.
 However, this possibility may not be realized in the type-B EWPT since $\lambda\gg\kappa$ is required.

In this paper, we investigate the feasibility of the EWBG in the UMSSM
in light of the recently discovered 126 GeV Higgs boson~\cite{126GeVHiggs}.
For the EWBG to be successful, a baryon number changing process in the broken phase 
has to be sufficiently suppressed.
This requirement enforces $v_C/T_C\gtrsim\zeta_{\rm sph}$, where
$T_C$ denotes a critical temperature, $v_C$ is a Higgs vacuum expectation value (VEV)
at $T_C$ and $\zeta_{\rm sph}$ is an $\mathcal{O}(1)$ value that depends on
the sphaleron energy and fluctuation determinants about the sphaleron.
To evaluate $v_C$ and $T_C$, the one-loop effective potential at finite temperature is used
\footnote{For gauge dependence issues in the perturbative analysis of the EWPT, see e.g. 
\cite{Patel:2011th,Wainwright:2012zn,Garny:2012cg} and references therein.}.
As for $\zeta_{\rm sph}$, it is usually set on 1 for simplicity in the literature. 
In our analysis, we explicitly work out $\zeta_{\rm sph}$ 
by taking a dominant contribution into account.
It is found that the strong first-order EWPT is driven by the doublet-singlet Higgs mixing effect.
Since the $Z'$ boson mass is mainly controlled by the singlet VEV (times $U(1)'$ coupling), 
the successful condition for the strong first-order EWPT inevitably leads to the light $Z'$ boson. 
Such a $Z'$ boson has to be leptophobic to be consistent with the collider bounds.

We also estimate the BAU using the closed-time-path (CTP) formalism.
Since the UMSSM contains an extra $CP$-violating phase coming from an interaction
between Higgsino and $Z'$-ino (superpartner of $U(1)'$ gauge boson ($Z'$)), 
we consider a scenario in which $Z'$-ino plays an essential role in generating the BAU.
Unlike the singlino-driven EWBG scenario in the NMSSM,  
it turns out that parameter space where the resonant enhancement of the BAU can occur
is compatible with the strong first-order EWPT.

The paper is organized as follows.
In Sec.~\ref{sec:Model}, we introduce the model and work out the Higgs boson masses
as well as the $Z'$ boson mass at the tree level.
The vacuum structure is also investigated.
In Sec.~\ref{sec:EWPT}, 
after introducing the one-loop effective potentials at zero and nonzero temperatures, 
we discuss characteristic features of the EWPT in the UMSSM.  
The sphaleron decoupling condition is studied in Sec.~\ref{sec:sphaleron}.
In Sec.~\ref{sec:BAU}, we briefly review the CTP formalism and 
derive the $CP$-conserving and -violating source terms in the quantum Boltzmann equation.
After that, we give the BAU formula. 
In Sec.~\ref{sec:numerics}, the numerical analysis is performed.
Before closing the section, the experimental constraints are discussed.
Sec.~\ref{sec:conclusion} is devoted to conclusions and discussion.

\section{The model}\label{sec:Model}
The UMSSM is one of the singlet-extended MSSM models, 
which may emerge as an effective theory of some unification model such as the $E_6$ model.
The gauge symmetry of the UMSSM 
is extended to $SU(3)_C\times SU(2)_L\times U(1)_Y\times U(1)'$,
where the extra $U(1)'$ is regarded as a remnant of larger gauge groups 
at an ultraviolet (UV) scale
(for a comprehensive review of supersymmetric $Z'$ models, see e.g. \cite{Langacker:2008yv}.).
The particle content of the UMSSM highly depends on the UV theory.
Usually, in addition to the MSSM particle content,  the exotic particles
are needed for the anomaly cancellation and the gauge coupling unification.
In this paper, we assume that such exotic particles are heavy enough not to affect
the electroweak scale phenomenology,
and concentrate on the subspace of the full theory. Specifically, the relevant
superpotential is
\begin{align}
W\ni \epsilon_{ij}(f_{AB}^{(e)}\widehat{H}_d^i\widehat{L}_A^j\widehat{E}_B
	+f_{AB}^{(d)}\widehat{H}_d^i\widehat{Q}_A^j\widehat{D}_B
	-f_{AB}^{(u)}\widehat{H}_u^i\widehat{Q}_A^j\widehat{U}_B
	-\lambda \widehat{S}\widehat{H}_d^i\widehat{H}_u^j),
\end{align}
where $\epsilon_{12}=-\epsilon_{21}=+1$, $f_{AB}^{(u,d,e)}$ denote Yukawa couplings 
with $A, B$ being family indices.
The $\widehat{S}^n~(n\in \bm{Z})$-type interactions are forbidden by the $U(1)'$ symmetry.

The Higgs potential at the tree level is given by the sum of $F$-, $D$- and soft-breaking terms
\begin{align}
V_0=V_F+V_D+V_{\rm soft},
\end{align}
where
\begin{align}
V_F&=|\lambda|^2\big\{|\epsilon_{ij}\Phi_d^i\Phi_u^j|^2+|S|^2(\Phi_d^\dagger\Phi_d
	+\Phi_u^\dagger\Phi_u)\big\},\\
V_D&=\frac{g_2^2+g_1^2}{8}(\Phi_d^\dagger\Phi_d-\Phi_u^\dagger\Phi_u)^2 
	+\frac{g_2^2}{2}(\Phi_d^\dagger\Phi_u)(\Phi_u^\dagger\Phi_d)\non\\
	&\quad+\frac{g'^2_1}{2}(Q_{H_d}\Phi_d^\dagger\Phi_d
	+Q_{H_u}\Phi_u^\dagger\Phi_u+Q_S|S|^2)^2,\\
V_{\rm soft}&=m_1^2\Phi_d^\dagger\Phi_d+m_2^2\Phi_u^\dagger\Phi_u 
	+m_S^2|S|^2-(\epsilon_{ij}\lambda A_{\lambda}S\Phi_d^i\Phi_u^j+{\rm h.c.}).
\end{align}
Here, $g_2$, $g_1$ and $g'_1$ are the gauge coupling constants of $SU(2)_L$, $U(1)_Y$ and $U(1)'$, respectively.
$Q_i~(i=H_d, H_u, S)$ denote the $U(1)'$ charges, which satisfies $Q_{H_d}+Q_{H_u}+Q_S=0$.
In this paper, we take $g_1' =\sqrt{5/3}g_1(\simeq 0.45)$ 
as motivated by the simple grand unified theories.
The Higgs fields are parametrized as
\begin{align}
\Phi_d&=
\left(
\begin{array}{c}
\frac{1}{\sqrt{2}}(v_d+h_d+ia_d) \\
\phi_d^-
\end{array}
\right),\quad 
\Phi_u=
e^{i\theta}\left(
\begin{array}{c}
\phi_u^+\\
\frac{1}{\sqrt{2}}(v_u+h_u+ia_u) 
\end{array}
\right), \\
S&=\frac{1}{\sqrt{2}}(v_S+h_S+ia_S),
\end{align}
where $\sqrt{v_d^2+v_u^2}\equiv v\simeq 246$ GeV, and we define $\tan\beta=v_u/v_d$.
The nonzero $\theta$ can break the $CP$ spontaneously.

The first derivatives of $V_0$ with respect to the fluctuation fields are, respectively, given by
\begin{align}
\frac{1}{v_d}\Vac{\frac{\del V_0}{\del h_d}}
&=m_{1}^{2}+\frac{g_{2}^{2}+g_{1}^{2}}{8}(v_{d}^{2}-v_{u}^{2})-R_\lambda\frac{v_uv_S}{v_d}
	+\frac{|\lambda|^2}{2}(v_u^2+v_S^2)+\frac{g'^2_1}{2}Q_{H_d}\Delta=0,\label{tad_hd}\\
\frac{1}{v_u}\Vac{\frac{\del V_0}{\del h_u}}
&=m_{2}^{2}-\frac{g_{2}^{2}+g_{1}^{2}}{8}(v_{d}^{2}-v_{u}^{2})-R_\lambda\frac{v_dv_S}{v_u}
	+\frac{|\lambda|^2}{2}(v_d^2+v_S^2)+\frac{g'^2_1}{2}Q_{H_u}\Delta=0,\label{tad_hu}\\
\frac{1}{v_S}\Vac{\frac{\del V_0}{\del h_S}}
&=m_S^2-R_\lambda\frac{v_dv_u}{v_S}
	+\frac{|\lambda|^2}{2}(v_d^2+v_u^2)+\frac{g'^2_1}{2}Q_{S}\Delta=0, \label{tad_hS}\\
\frac{1}{v_u}\Vac{\frac{\del V_0}{\del a_d}}
&=\frac{1}{v_d}\Vac{\frac{\del V_0}{\del a_u}}=I_\lambda v_S=0,\label{tad_ad}\\
\frac{1}{v_S}\Vac{\frac{\del V_0}{\del a_S}}
&=I_\lambda \frac{v_dv_u}{v_S}=0,\label{tad_aS}
\end{align}
where $\langle X\rangle$ denotes that $X$ should be evaluated in the vacuum,
and $v_{d,u,S}$ are assumed to be nonzero. The symbols $\Delta$, $R_\lambda$ and $I_\lambda$ are defined by
\begin{align}
\Delta &= Q_{H_d}v_d^2+Q_{H_u}v_u^2+Q_{S}v_S^2,\\
R_\lambda
&=\frac{|\lambda| |A_\lambda|}{\sqrt{2}}
	\cos(\delta_{A_\lambda}+\delta'_\lambda), \quad
I_\lambda
= \frac{|\lambda| |A_\lambda|}{\sqrt{2}}
	\sin(\delta_{A_\lambda}+\delta'_\lambda),
\end{align}
with $\delta'_\lambda=\delta_\lambda+\theta$.
In our analysis, $m_{1,2,S}^2$ are fixed by Eqs.~(\ref{tad_hd}), (\ref{tad_hu}) and (\ref{tad_hS}).
$I_\lambda=0$ enforces no $CP$ violation at the tree level.
At this point, there are twofold ambiguities about the sign of $R_\lambda$.
However, it turns out that the sign of $R_\lambda$ has to be taken positive to be consistent
with the physical $CP$-odd Higgs boson mass.
%
%
\subsection{Higgs and $Z'$ boson masses}\label{subsec:mH_mZp}
In our analysis, the Higgs boson masses are calculated to the one-loop level using the 
effective potential. In this section, we give approximate mass formulas.

The lightest Higgs boson mass is bounded from above as
\begin{align}
m_{H_1}^2\leq m_Z^2\cos^22\beta+\frac{|\lambda|^2}{2}v^2\sin^22\beta
	+g'^2_1v^2(Q_{H_d}\cos^2\beta+Q_{H_u}\sin^2\beta)^2.\label{mh_MAX}
\end{align}
In addition to the ordinary MSSM-like $D$-term contribution, 
the $F$ and $D'$-term contributions also show up in the UMSSM,
so the $m_{H_1}=126$ GeV is easily realized even at the tree level.
For instance, the right-hand side of Eq.~(\ref{mh_MAX}) would be around $(126~{\rm GeV)^2}$
for $\tan\beta=1$, $|\lambda|=0.72$ and $Q_{H_d}=Q_{H_u}=-1/2$.
In this case, the most contributions come from the $F$-term.
More accurate expression of $m_{H_1}$ 
can be derived by taking the Higgs doublet-singlet mixing into account.
The three $CP$-even Higgs boson masses are found to be
\begin{align}
m_{H_{1,2}}^2 &= \frac{1}{2}
\bigg[
	m_S^2+|\lambda|^2v^2+6g'^2_1Q^2v_S^2 \non\\
&\hspace{1cm}	
	\mp\sqrt{\big\{m_S^2+2g'^2_1Q^2(3v_S^2-v^2)\big\}^2
	+4v^2\big\{R_\lambda-(|\lambda|^2-2g'^2_1Q^2)v_S\big\}^2}
\bigg], \label{mH12_tree}\\
m_{H_3}^2 & = m_Z^2-\frac{|\lambda|^2}{2}v^2+2R_\lambda v_S,
\end{align}
where $Q_{H_d}=Q_{H_u}\equiv Q$ and $\tan\beta=1$ are taken for an illustration.
Note that $m_{H_1}^2$ could be negative if $R_\lambda$ is sufficiently large,
which is the common feature in the singlet-extended MSSMs (see e.g. \cite{Miller:2003ay}).
Such a tachyonic mass can be evaded if $R_\lambda\simeq(|\lambda|^2-2g'^2_1Q^2)v_S$.

After straightforward calculation, one can obtain the tree-level $CP$-odd Higgs boson mass
\begin{align}
m^2_A = \frac{2R_\lambda v_S}{\sin2\beta}\left(1+\frac{v^2}{4v^2_S}\sin^22\beta\right).
\end{align}
As mentioned above, the sign of $R_\lambda$ has to be positive in order for $m_A^2>0$.

The tree-level charged Higgs boson mass is calculated as
\begin{align}
m_{H^\pm}^2=\frac{1}{\sin\beta\cos\beta}\Vac{\frac{\del^2 V_0}{\del\phi_d^+\del\phi_u^-}}
	=m_W^2+\frac{2R_\lambda}{\sin2\beta}v_S-\frac{|\lambda|^2}{2}v^2.\label{mch_tree}
\end{align}
In the MSSM limit in which $\lambda\to 0$ and $v_S\to \infty$ with $\lambda v_S$ fixed,
we can recover the mass relationship of $m_{H^\pm}^2=m_W^2+m_A^2$.
In our investigation, we will trade $|A_\lambda|$ with the one-loop-corrected charged Higgs boson mass
and take $m_{H^\pm}$ as the input parameter.

Because of the presence of the $Z'$ boson, the ordinary $Z$ boson can mix with it.
Consequently, the mass matrix of the neutral $Z$ bosons takes the 2-by-2 form
\begin{align}
\mathcal{M}^2_{ZZ'}=\left(
\begin{array}{cc}
m^2_Z & \frac{g'_1}{2}\sqrt{g_2^2+g_1^2}(Q_{H_d}v_d^2-Q_{H_u}v_u^2) \\
\frac{g'_1}{2}\sqrt{g_2^2+g_1^2}(Q_{H_d}v_d^2-Q_{H_u}v_u^2) & m^2_{Z'}
\end{array}
\right),
\end{align}
where
\begin{align}
m^2_Z =\frac{g^2_2+g^2_1}{4}v^2,\quad
m^2_{Z'} =g'^2_1(Q_{H_d}^2v_d^2+Q_{H_u}^2v_u^2+Q_S^2v_S^2). 
\label{mZ-mZp}
\end{align}
The eigenvalues of the mass matrix and the mixing angle are, respectively, given by
\begin{align}
m^2_{Z_{1,2}} &=\frac{1}{2}\left[m^2_Z
	+m^2_{Z'}\pm\sqrt{(m^2_Z-m^2_{Z'})^2
	+g'^2_1(g^2_2+g^2_1)(Q_{H_d}v^2_d-Q_{H_u}v^2_u)^2} \right], \\
\alpha^{}_{ZZ'}&=\frac{1}{2}\arctan\left(\frac{2(\mathcal{M}^2_{ZZ'})_{12}}
	{(\mathcal{M}^2_{ZZ'})_{22}-(\mathcal{M}^2_{ZZ'})_{11}} \right).
\end{align}
Since the electroweak precision tests impose $\alpha_{ZZ'}<\mathcal{O}(10^{-3})$,
we will take $(\mathcal{M}_{ZZ'}^2)_{12}=0$, leading to $\tan\beta=\sqrt{|Q_{H_d}|/|Q_{H_u}|}$
\footnote{Note that since the $Z$-$Z'$ mixing arises at the loop level, the tree-level relation would be modified.
Therefore, we may choose the parameters more judiciously. 
This point will be discussed in \ref{subsec:exp}.}. 
Thus, the masses of the $Z$ and $Z'$ bosons are simply given by Eq.~(\ref{mZ-mZp}).
%
%
\subsection{Vacuum structures}\label{subsec:Veff1}
We use the effective potential to study the EWPT.
First, we parametrize the classical background fields as
\begin{align}
\langle \Phi_d \rangle = 
\left(
	\begin{array}{c}
	\frac{1}{\sqrt{2}}\varphi_d \\
	0 
	\end{array}
\right),\quad
\langle \Phi_u \rangle = 
\left(
	\begin{array}{c}
	0 \\
	\frac{e^{i\vartheta}}{\sqrt{2}}\varphi_u
	\end{array}
\right),\quad
\langle S \rangle = \frac{1}{\sqrt{2}}\varphi_S.
\end{align}
In the vacuum, $\varphi_d=v_d$, $\varphi_u=v_u$, $\vartheta=\theta$ and $\varphi_S=v_S$.
The tree-level effective potential is 
\begin{align}
V_0(\varphi_d, \varphi_u, \vartheta, \varphi_S)
&=\frac{1}{2}m_1^2\varphi_d^2+\frac{1}{2}m_2^2\varphi_u^2
	+\frac{1}{2}m_S^2\varphi_S^2-R_\lambda \varphi_d\varphi_u\varphi_S
	+\frac{g_2^2+g_1^2}{32}(\varphi_d^2-\varphi_u^2)^2 \non\\
&\quad+\frac{|\lambda|^2}{4}(\varphi_d^2\varphi_u^2+\varphi_d^2\varphi_S^2+\varphi_u^2\varphi_S^2)
	+\frac{g'^2_1}{8}(Q_{H_d}\varphi_d^2+Q_{H_u}\varphi_u^2+Q_{S}\varphi_S^2)^2.	
	\label{V0}
\end{align}
Before going to the EWPT study, we briefly discuss vacuum structure at the tree level.
Because of the additional singlet Higgs field, the Higgs potential can have various vacua.
We require that the EW phase should be the global minimum,
\begin{align}
V_0(\boldsymbol{\varphi}=\boldsymbol{v}_{\rm EW})
<V_0(\boldsymbol{\varphi}\neq\boldsymbol{v}_{\rm EW}),
\end{align}
where we categorize the diverse vacua as~\cite{Funakubo:2005pu}
\begin{align}
{\rm EW}: v=246~{\rm GeV},~v_S\neq0;\quad
{\rm I} : v=0,~v_S\neq 0; \quad
{\rm II} : v\neq0,~v_S= 0; \quad
{\rm SYM} : v=v_S=0.
\end{align}
The energy difference of the EW and SYM phases is
\begin{align}
\Delta^{(\rm SYM-EW)}V_0 
&\equiv V_0^{(\rm SYM)}(0, 0, 0, 0)-V_0^{(\rm EW)}(v_d, v_u, \theta, v_S) \non\\
&= \frac{g_2^2+g_1^2}{32}(v_d^2-v_u^2)^2-\frac{1}{2}R_\lambda v_dv_uv_S
	+\frac{|\lambda|^2}{4}(v_d^2v_u^2+v_d^2v_S^2+v_u^2v_S^2)
	+\frac{g'^2_1}{8}\Delta^2,
\end{align}
We can see that $\Delta^{(\rm SYM-EW)}V_0$ could be negative for a sufficiently large $R_\lambda$.
Conversely, $\Delta^{(\rm SYM-EW)}V_0>0$ yields the upper bound of $m_{H^\pm}^2$ as
\begin{align}
m_{H^\pm}^2 
	< m_W^2+m_Z^2\cot^22\beta+\frac{2|\lambda|^2v_S^2}{\sin^22\beta}
		+\frac{g'^2_1\Delta^2}{v^2\sin^22\beta}
	\equiv(m_{H^\pm}^{\rm max})^2,\label{mch_max}
\end{align}
where we use Eq.~(\ref{mch_tree}) to replace $R_\lambda$ with $m_{H^\pm}$.
\begin{figure}[t]
\center
\includegraphics[width=7.5cm]{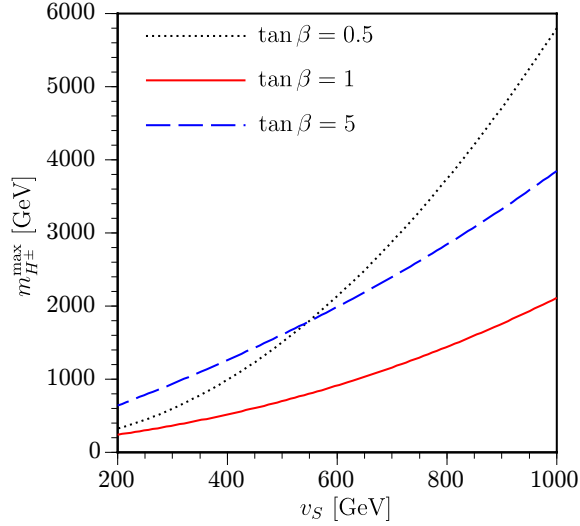}
\caption{$m_{H^\pm}^{\rm max}$ is plotted as a function of $v_S$ varying $\tan\beta=0.5, 1, 5$.
We take $|\lambda|=0.8$, $Q_{H_d}=-0.5$ and $Q_{H_u}=Q_{H_d}/\tan^2\beta$.}
\label{fig:mchMAX_vS}
\end{figure}
In Fig.~\ref{fig:mchMAX_vS}, $m_{H^\pm}^{\rm max}$ is shown for a few sample points.
Here, we set $|\lambda|=0.8$, $Q_{H_d}=-0.5$ and $Q_{H_u}=Q_{H_d}/\tan^2\beta$, with $\tan\beta=0.5, 1, 5$.
Since the dominant terms are proportional to $1/\sin^22\beta$ in Eq.~(\ref{mch_max}), 
the smallest $m_{H^\pm}^{\rm max}$ is realized for $\tan\beta=1$.
In this case, $m_{H^\pm}^{\rm max}$ cannot exceed about 1 TeV if $v_S\lesssim 640$ GeV.
Depending on the theory parameters, $\Delta^{(\rm I-EW)}V_0>0$ can yield a more severe upper bound of $m_{H^\pm}$.
At the one-loop level, it is not easy to obtain the analytic expressions for the global minimum condition.
Therefore, we will search for a global minimum numerically.

\section{Electroweak phase transition}\label{sec:EWPT}
The ${\rm \overline{DR}}$-regularized one-loop effective potential at zero temperature is~\cite{Coleman:1973jx}
\begin{align}
V_1(\varphi_d, \varphi_u, \vartheta, \varphi_S) = \sum_{i}n_i
	\frac{\bar{m}_i^4}{64\pi^2}\left(\ln\frac{\bar{m}_i^2}{\bar{\mu}^2}-\frac{3}{2}\right),
	\label{V1zero}
\end{align}
where $\bar{m}_i^2$ denote the field-dependent masses with $i=W^\pm, Z, Z', t,b, \tilde{t}_{1,2}, \tilde{b}_{1,2}$,
and $\bar{\mu}$ is a renormalization scale and
$n_i$ is respectively given by
\begin{align}
n_W&=6, \quad n_Z =n_{Z'}=3,\quad n_t=n_b=-4N_C, \quad 
n_{\tilde{t}_{1,2}}=n_{\tilde{b}_{1,2}}=2N_C,
\end{align}
with $N_C$ being the color degree of freedom.

The global minimum search at zero temperature
is performed using Eqs.~(\ref{V0}) and (\ref{V1zero}).
The one-loop effective potential at nonzero temperatures is
\begin{align}
V_1(\varphi_d, \varphi_u, \vartheta, \varphi_S; T) = \sum_{i}n_i
	\frac{T^4}{2\pi^2}I_{B,F}\left(\frac{\bar{m}_i^2}{T^2}\right),
\end{align}
where $I_{B,F}(a^2)$ are defined by
\begin{align}
I_{B,F}(a^2) &= \int_0^\infty dx~x^2\ln\Big(1\mp e^{-\sqrt{x^2+a^2}}\Big).
\end{align}
In our numerical analysis, we use the fitting functions of $I_{B,F}(a^2)$ 
that are employed in Ref.~\cite{Funakubo:2009eg}.
The relative errors of them are less than $10^{-6}$ which is sufficient for our study.

As is well known, a naive perturbative expansion in a coupling constant 
can be invalidated by sizable thermal loop corrections at high temperatures.
To obtain meaningful results, such corrections have to be resummed in a consistent way.
As for the dominant corrections (daisy diagrams), 
the resultant resummation formula has the compact form
\begin{align}
V_{\rm daisy}(\varphi_d, \varphi_u, \vartheta, \varphi_S; T)  
&= -\sum_{j={\rm squarks}}n_j\frac{T}{12\pi}
\Big[(\bar{M}_j^2)^{3/2}-(\bar{m}_j^2)^{3/2}\Big] \non\\
&\quad -\sum_{j={\rm gauge}}n_j\frac{T}{12\pi}
\Big[(\bar{M}_{Lj}^2)^{3/2}-(\bar{m}_j^2)^{3/2}\Big],\label{Vdaisy}
\end{align}
where $\bar{M}^2 = \bar{m}^2+\Sigma(T)$ with $\Sigma(T)$ being the thermal self-energy
of the squarks and $\bar{M}_L^2 = \bar{m}^2+\Pi(T)$,
where $\Pi(T)$ is defined by the longitudinal part of the gauge boson self-energy 
in the infrared limit, specifically,
$\Pi(T)=\lim_{p^0=0, \boldsymbol{p}\to 0}\Pi_{00}(p^0, \boldsymbol{p}; T)$.
In our analysis, since the squarks are heavy enough to decouple from the thermal bath,
the only gauge boson contributions are taken into account in Eq.~(\ref{Vdaisy}). 
$\Pi(T)$ is evaluated to leading order in the high-temperature expansion, 
so typically $\Pi(T)\simeq \mathcal{O}(T^2)$.
The explicit formulas of $\Pi(T)$ are presented in Appendix~\ref{app:mT}.
It turns out that since the first-order EWPT is mainly driven by the tree-level effect,
the daisy resummation does not alter the results by more than a few \%.

In order to get a first-order EWPT, there must be a negative contribution in $V_{\rm eff}$,
which induces a barrier between the two degenerate minima.
In the SM or the MSSM, such a negative contribution comes from the bosonic thermal loops,
and thus the effects are loop suppressed.
In the singlet-extended models, on the other hand, the mixing terms between singlet and doublet Higgs fields
that exist in the tree-level potential can generate the negative contributions 
which may drive the first-order EWPT.
It is easy to understand this mechanism in a simplified potential in which the tree-level potential with $g'_1=0$ 
and the $\mathcal{O}(T^2)$ corrections are taken into account~\cite{Menon:2004wv,Huber:2006wf,Chiang:2009fs,Carena:2011jy}.
Specifically, 
\begin{align}
V_{\rm eff}(\varphi; T)=\frac{1}{2}M^2(T)\varphi^2+\frac{1}{2}m_S^2\varphi_S^2
	-\tilde{R}_\lambda\varphi^2\varphi_S+\frac{|\lambda|^2}{4}\varphi^2\varphi_S^2
	+\frac{\tilde{\lambda}^2}{4}\varphi^4, \label{Veff_sim}
\end{align}
where
\begin{align}
M^2(T) &= m^2_1\cos^2\beta+m^2_2\sin^2\beta+\mathcal{G} T^2, \label{MT}\\
\tilde{R}_\lambda & = R_\lambda\sin\beta\cos\beta,
\quad \tilde{\lambda}^2 = \frac{g^2_2+g^2_1}{8}\cos^22\beta+\frac{|\lambda|^2}{4}\sin^22\beta,
\end{align}
The last term in Eq.~(\ref{MT}) denotes the dominant thermal correction with
$\mathcal{G}$ being the sum of the relevant couplings.
Here, the temperature dependence of $\beta$ is neglected for the sake of simplicity. 
After eliminating $\varphi_S$ using the minimization condition with respect to $\varphi_S$, 
Eq.~(\ref{Veff_sim}) is reduced to
\begin{align}
V_{\rm eff}(\varphi; T)=\frac{1}{2}M^2(T)\varphi^2
	-\frac{\tilde{R}_\lambda^2\varphi^4}{2(m_S^2+|\lambda|^2\varphi^2/2)}
	+\frac{\tilde{\lambda}^2}{4}\varphi^4.
\end{align}
The second term can be the source of the negative contribution.
The first-order EWPT may be realized if $\tilde{\lambda}<\sqrt{2/m_S^2}|\tilde{R}_\lambda|$~\cite{Menon:2004wv,Huber:2006wf,Chiang:2009fs}.

In Sec.~\ref{sec:numerics}, using the daisy-improved one-loop effective potential
we will evaluate the critical temperature $(T_C)$ which
is defined by the temperature at which $V_{\rm eff}$ has the two degenerate minima,
and the Higgs VEVs at $T_C$ denoted as 
\begin{align}
v_C &=\lim_{T\uparrow T_C}\sqrt{v_d^2(T_C)+v_u^2(T_C)},\quad
\theta_C = \lim_{T\uparrow T_C}\theta(T_C),\quad 
v_{SC}=\lim_{T\uparrow T_C}v_S(T_C),\\
v_C^{\rm sym} &=\lim_{T\downarrow T_C}\sqrt{v_d^2(T_C)+v_u^2(T_C)}=0,\quad
\theta_C^{\rm sym} = \lim_{T\downarrow T_C}\theta(T_C),\quad 
v_{SC}^{\rm sym}=\lim_{T\downarrow T_C}v_S(T_C).
\end{align}
In the parameter space we search for in this analysis, it turns out that 
the spontaneous $CP$ violation does not occur, and thus we have $\theta_C=\theta_C^{\rm sym}=0$
As noted in Ref.~\cite{Chiang:2009fs}, 
$|v_{SC}-v_{SC}^{\rm sym}|$ may be greater than some value
if the EWPT is strongly first-order.

\section{Sphaleron decoupling condition}\label{sec:sphaleron}
For the EWBG to be successful, the $B$-changing rate in the broken phase ($\Gamma_B^{(b)}(T)$) 
has to be sufficiently suppressed in order not to erase the generated BAU.
Comparing $\Gamma_B^{(b)}(T)$ with the Hubble constant ($H(T)$), one gets
\begin{align}
\Gamma_B^{(b)}(T)\simeq ({\rm prefactor})e^{-E_{\rm sph}(T)/T} 
< H(T)\simeq 1.66\sqrt{g_*(T)}T^2/m_{\rm P}, \label{GamvsH}
\end{align}
where $E_{\rm sph}$ denotes the sphaleron energy, 
(prefactor) includes the fluctuation determinants around the sphaleron etc~\cite{Arnold:1987mh,Funakubo:2009eg},
$g_*$ is the degrees of freedom of relativistic particles in the plasma
($g_*=106.75$ in the SM) and $m_{\rm P}$ stands for the Planck mass 
which is about $1.22\times 10^{19}$ GeV.
Since $E_{\rm sph}$ is proportional to the Higgs VEV, 
Eq.~(\ref{GamvsH}) would be satisfied if the EWPT is strongly first-order.
Conventionally, the sphaleron energy is parametrized as $E_{\rm sph}(T)=4\pi v(T)\mathcal{E}(T)/g_2$.
Eq.~(\ref{GamvsH}) is then cast into the form
\begin{align}
\frac{v(T)}{T} > \frac{g_2}{4\pi \mathcal{E}(T)}
\Big[
	42.97+\mbox{log corrections}
\Big]\equiv \zeta_{\rm sph}.\label{sph_dec}
\end{align}
The dominant contributions on the right-hand side is $\mathcal{E}(T)$
while the log corrections that mostly come from the zero mode factors of the fluctuations 
about the sphaleron typically amount to about 10\%~\cite{Funakubo:2009eg}.
Note that the decoupling condition (\ref{sph_dec}) should be evaluated at a temperature 
such that the EWPT completes. 
However, since it is difficult to estimate such a temperature,
we evaluate Eq.~(\ref{sph_dec}) at $T_C$.
This simplification would be justified as long as supercooling is small.

We may evaluate $\mathcal{E}$ at the zero temperature neglecting its temperature dependence.
For the SM Higgs boson with a mass of 126 GeV, $\mathcal{E}(0)\simeq1.92$ is found
at the tree level without the $U(1)_Y$ contributions~\cite{sph_SM,Senaha:2013fva}.
With this $\mathcal{E}(0)$, one obtains $\zeta_{\rm sph}=1.16$,
where only the dominant contributions are retained on the right-hand side in Eq.~(\ref{sph_dec}).
It should be noted that the use of $\mathcal{E}(0)$ in the decoupling criterion 
leads to somewhat underestimated results since $\mathcal{E}(T)<\mathcal{E}(0)$. 
In the MSSM, using the finite-temperature effective potential at the one-loop level, 
$v(T_N)/T_N>1.38$ is obtained, where the sphaleron energy as well as the translational 
and rotational zero mode factors of the fluctuation around the sphaleron 
are evaluated at a nucleation temperature ($T_N$) which is somewhat below 
$T_C$~\cite{Funakubo:2009eg}.
Note that the common choice of $\zeta_{\rm sph}=1$ or 0.9 in the literature 
may lead to the underestimated sphaleron decoupling condition.

To find the sphaleron solutions in the UMSSM, 
we adopt a spherically symmetric configuration ansatz
with a noncontractible loop as in the the SM without the $U(1)_Y$ corrections~\cite{sph_SM}. 
To do so, we take $g_1=g'_1=0$.
Similar analysis in the NMSSM can be found in~\cite{Funakubo:2005bu}.
Actually, the sphalerons in our case may correspond to those in the NMSSM 
with a vanishing trilinear coupling of the singlet Higgs field.

In the presence of the magnetic field, the sphaleron energy would be lowered due to the
sphaleron magnetic dipole moment (see e.g. \cite{DeSimone:2011ek}). 
The detailed studies of such an effect will be performed elsewhere.

The noncontractible loop configurations for $A_i$, $\Phi_{d,u}$ and $S$ are, respectively, given by
\begin{align}
A_i(\mu, r, \theta, \phi) &= \frac{i}{g_2}f(r)\partial_i U(\mu, \theta, \phi)U^{-1}(\mu, \theta, \phi), \\
\Phi_d(\mu, r, \theta, \phi) &= \frac{v_d}{\sqrt{2}}
	\left\{
	(1-h_1(r))\left(
			\begin{array}{c}
			e^{i\mu}\cos\mu \\
			0
			\end{array}
			\right)
	+h_1(r)\tilde{\Phi}_d(\mu, \theta, \phi)
	\right\}, \\
\Phi_u(\mu, r, \theta, \phi) &= \frac{v_u}{\sqrt{2}}
	\left\{
	(1-h_2(r))\left(
			\begin{array}{c}
			0 \\
			e^{-i\mu}\cos\mu
			\end{array}
			\right)
	+h_2(r)\tilde{\Phi}_u(\mu, \theta, \phi)
	\right\}, \\
S(\mu, r, \theta, \phi) &= \frac{v_S}{\sqrt{2}}k(r),
\end{align}
where
\begin{align}
U(\mu, \theta, \phi) = 
	\left(
		\begin{array}{cc}
		e^{i\mu}(\cos\mu-i\sin\mu\cos\theta) & e^{i\phi}\sin\mu\sin\theta \\
		-e^{-i\phi}\sin\mu\sin\theta & e^{-i\mu}(\cos\mu+i\sin\mu\cos\theta)
		\end{array}
	\right),
\end{align}
which is noncontractible since $\pi_3(SU(2))\simeq \boldsymbol{Z}$.
$(r, \theta, \phi)$ represents the spherical polar coordinates, and $\mu$ is the loop parameter
running from 0 to $\pi$,
which parameterizes the least energy path between two adjacent topologically distinct vacua.
So the $\mu=0$ and $\pi$ correspond to the vacuum configurations 
while $\mu=\pi/2$ yields the saddle point configuration, {\it i.e.,} sphaleron.
$\tilde{\Phi}_d(\mu, \theta, \phi)$ and $\tilde{\Phi}_u(\mu, \theta, \phi)$ are 
\begin{align}
\tilde{\Phi}_d(\mu, \theta, \phi) &= U(\mu, \theta, \phi)
	\left(
		\begin{array}{c}
		1 \\
		0
		\end{array}
	\right)
	=
	\left(
		\begin{array}{c}
		e^{i\mu}(\cos\mu-i\sin\mu\cos\theta) \\
		-e^{-i\phi}\sin\mu\sin\theta
		\end{array}
	\right), \\
\tilde{\Phi}_u(\mu, \theta, \phi) &= U(\mu, \theta, \phi)
	\left(
		\begin{array}{c}
		0 \\
		1
		\end{array}
	\right)
	=
	\left(
		\begin{array}{c}
		e^{i\phi}\sin\mu\sin\theta \\
		e^{-i\mu}(\cos\mu+i\sin\mu\cos\theta)
		\end{array}
	\right).
\end{align}
The energy functional at $\mu=\pi/2$ is reduced to
\begin{align}
E_{\rm sph}[f, h_1, h_2, k] &= \frac{4\pi \Omega}{g_2}\int_0^\infty d\xi
	\bigg[
		4f'^2+\frac{8}{\xi^2}(f-f^2)^2
		+\frac{v_d^2}{2\Omega^2}\Big\{\xi^2h_1'^2+2h_1^2(1-f)^2\Big\}\non\\
&\hspace{3cm}		
		+\frac{v_u^2}{2\Omega^2}\Big\{\xi^2h_2'^2+2h_2^2(1-f)^2\Big\}		
		+\frac{\xi^2v_S^2}{2\Omega^2}k'^2
		+\frac{\xi^2}{g_2^2\Omega^4}V_0(h_1, h_2, k)
	\bigg], \label{Esph}
\end{align}
where $\xi=g_2\Omega r$ with $\Omega$ being an arbitrary parameter with a mass dimension,
and the prime on the profile functions ($f,h_{1,2},k$) denotes the derivative with respect to $\xi$.
As mentioned above, $\Omega$ is usually set on $v$.
The explicit form of $V_0(h_1, h_2, k)$ is 
\begin{align}
V_0(h_1, h_2, k) &= \frac{1}{2}v_dv_uv_SR_\lambda(h_1^2+h_2^2+k^2-2h_1h_2k-1) 
	+\frac{g_2^2}{32}\Big[v_d^2(h_1^2-1)-v_u^2(h_2^2-1)\Big]^2  \non\\
&\quad
	+\frac{|\lambda|^2}{4}
	\Big[
		v_d^2v_u^2(-h_1^2-h_2^2+h_1^2h_2^2+1)
		+v_S^2\big\{v_d^2(h_1^2-1)+v_u^2(h_2^2-1)\big\}(k^2-1)		
	\Big].
\end{align}
With Eq.~(\ref{Esph}), the equations of motion for the sphaleron are
\begin{align}
\frac{d^2f}{d\xi^2} &= \frac{2}{\xi^2}f(1-f)(1-2f)
	-\frac{1}{4\Omega^2}(v_d^2h_1^2+v_u^2h_2^2)(1-f), \\
\frac{d}{d\xi}\left(\xi^2\frac{dh_1}{d\xi}\right)
&= 2h_1(1-f)^2+\frac{\xi^2}{g_2^2v_d^2\Omega^2}\frac{\partial V_0}{\partial h_1}, \\
\frac{d}{d\xi}\left(\xi^2\frac{dh_2}{d\xi}\right)
&= 2h_2(1-f)^2+\frac{\xi^2}{g_2^2v_u^2\Omega^2}\frac{\partial V_0}{\partial h_2}, \\
\frac{d}{d\xi}\left(\xi^2\frac{dk}{d\xi}\right)
&=\frac{\xi^2}{g_2^2v_S^2\Omega^2}\frac{\partial V_0}{\partial k}.
\end{align}
We solve these equations under following boundary conditions
\begin{align}
&\lim_{\xi\to 0}f(\xi)=\lim_{\xi\to 0}h_1(\xi)=\lim_{\xi\to 0}h_2(\xi)=\lim_{\xi\to 0}k'(\xi)=0,\\
&\lim_{\xi\to\infty}f(\xi)=\lim_{\xi\to\infty}h_1(\xi)=\lim_{\xi\to\infty}h_2(\xi)
	=\lim_{\xi\to\infty}k(\xi)=1.
\end{align}
Note that the Neumann boundary condition is imposed for $k(\xi)$ at $\xi=0$~\cite{Funakubo:2005bu}.

\section{Baryon asymmetry}\label{sec:BAU}
We start by giving a brief review on the CTP formalism to make the paper self-contained.
We closely follow Refs.~\cite{CTP_Riotto,Lee:2004we}.
The closed time path is defined as the path from $-\infty$ to $+\infty$ and back to $-\infty$.
Accordingly, the fermion propagator in the CTP formalism has the 2-by-2 form
\begin{align}
\hat{S}(x, y)= \langle T_P\psi(x)\bar{\psi}(y)\rangle=
\left(
\begin{array}{cc}
S^t(x, y) & S^<(x, y) \\
S^>(x, y) & S^{\bar{t}}(x, y)
\end{array}
\right),
\end{align}
where $T_P$ denotes the path-ordering symbol, and each component is defined by
\begin{align}
S^>(x, y) &= \langle\psi(x)\bar{\psi}(y)\rangle,\\
S^<(x, y) &= -\langle\bar{\psi}(y)\psi(x)\rangle,\\
S^t(x, y) &= \langle T\{\psi(x)\bar{\psi}(y)\}\rangle
	= \theta(x^0-y^0)S^>(x, y)+\theta(y^0-x^0)S^<(x, y),\\
S^{\bar{t}}(x, y) &= \langle\bar{T}\{\psi(x)\bar{\psi}(y)\}\rangle
	= \theta(x^0-y^0)S^<(x, y)+\theta(y^0-x^0)S^>(x, y),
\end{align}
where $S^{t(\bar{t})}(x,y)$ are (anti-) time-ordered propagators.
The self-energy in the CTP formalism takes the form
\begin{align}
\hat{\Sigma}(x, y)=
\left(
\begin{array}{cc}
\Sigma^t(x, y) & \Sigma^<(x, y) \\
\Sigma^>(x, y) & \Sigma^{\bar{t}}(x, y)
\end{array}
\right).
\end{align}
The boson propagator and self-energy are also defined in the same manner.
The divergence of the Noether current can be expressed in terms of $S^{>}(x, y)$ or $S^{<}(x, y)$ as
\begin{align}
\partial_\mu j_\psi^\mu(x)
&=i\cdot{\rm tr}\Big\{
	i\gamma_\mu(\rpartialupx+\lpartialupy)S^{>,<}(x, y)\Big\}\Big|_{x=y}.\label{divJ}
\end{align}
$\hat{S}(x, y)$ satisfies the Schwinger-Dyson equation which is given by
\begin{align}
\hat{S}(x, y)&=\hat{S}^0(x, y)-i\int_Cd^4z\int_Cd^4w~\hat{S}^0(x, z)\hat{\Sigma}(z, w)
	\hat{S}(w, y),\label{SD1}
\end{align}
or
\begin{align}
\hat{S}(x, y)&=\hat{S}^0(x, y)-i\int_Cd^4z\int_Cd^4w~\hat{S}(x, z)\hat{\Sigma}(z, w)
	\hat{S}^0(w, y),\label{SD2}
\end{align}
where $\hat{S}^0(x, y)$ denotes the noninteracting Green's function and
$C$ represents the closed time path defined above. 
With Eqs.~(\ref{divJ}), (\ref{SD1}) and (\ref{SD2}), it follows that
\begin{align}
\frac{\partial n_\psi(X)}{\partial t_X}+\nabla_{\scriptsize{\bm{X}}}\cdot\bm{j}_\psi(X)
&= i\int_{-\infty}^{t_X} dz^0\int_{-\infty}^{\infty}d^3\bm{z}~
	{\rm tr}\Big[\Sigma^>(X, z)S^<(z, X)-\Sigma^<(X, z)S^>(z, X) \non\\
	&\quad\hspace{2.5cm}-S^>(X, z)\Sigma^<(z, X)+S^<(X, z)\Sigma^>(z, X)\Big],\label{QDE}
\end{align}
where $X^\mu=(t_X, \boldsymbol{X})=(x^\mu+y^\mu)/2$.

We work out the right hand side of Eq.~(\ref{QDE}) to derive the $CP$-violating source terms
and the chirality-flipping rate via the Higgs bubble walls. 
Here, we focus only on the contributions of the $Z'$-ino to the neutral Higgsino current ($j_{\widetilde{H}^0}$). 
The relevant neutral Higgsino-$Z'$-ino interactions are given by
\begin{align}
\mathcal{L}_{\widetilde{H}^0\widetilde{Z}'} 
&=\overline{\widetilde{H}^0}(i\gamma^\mu\partial_\mu-|\mu_{\rm eff}(x)|)\widetilde{H}^0
+\frac{1}{2}\overline{\widetilde{Z}'}(i\gamma^\mu\partial_\mu-|M'_1|)\widetilde{Z}' \non\\
&\quad-g'_1
\bigg[	
	\overline{\widetilde{H}^0}
	(Q_{H_d}v_d(x)e^{-i\delta_{M'_1}/2}P_L
		-Q_{H_u}v_u(x)e^{i(\delta_\lambda+\theta(x)+\delta_{M'_1}/2)}P_R
	)\widetilde{Z}' \non\\
&\hspace{2cm}+\overline{\widetilde{Z}'}
	(-Q_{H_u}v_u(x)e^{-i(\delta_\lambda+\theta(x)+\delta_{M'_1}/2)}P_L
		+Q_{H_d}v_d(x)e^{i\delta_{M'_1}/2}P_R
	)\widetilde{H}^0
\bigg],
\end{align}
where $\mu_{\rm eff}(x)=\lambda v_S(x)e^{i\theta(x)}/\sqrt{2}$, $M'_1=|M_1'|e^{i\delta_{M_1'}}$
and $P_{L,R}=(1\mp \gamma_5)/2$ with $\gamma_5={\rm diag}(-1,1)$.
As mentioned in Sec.~\ref{sec:EWPT}, we do not consider the parameter space 
where the spontaneous $CP$ violation occurs
and hence $\theta(x)=0$.
The $CP$-violating source is originated from the imaginary part of the right hand side in Eq.~(\ref{QDE}). 
Using the VEV insertion method and the derivative expansion with respect to the bubble walls, 
the $CP$-violating source term may be cast into the form
\begin{align}
S_{\widetilde{H}^0}(X)
\ni -4g'^2_1Q_{H_d}Q_{H_u}|M'_1||\mu_{\rm eff}(X)|v^2(X)\del_{t_X}\beta(X)
	\sin(\delta_\lambda+\delta_{M'_1})\mathcal{I}_{\widetilde{Z}'\widetilde{H}^0}^f,
\end{align}
where $\mathcal{I}_{\widetilde{Z}'\widetilde{H}^0}^f$ is given in Appendix \ref{app:Tfunc}.
Likewise, the $CP$-conserving chirality-flipping rate via the Higgs bubble walls is calculated from 
the real part of the right hand side in Eq.~(\ref{QDE}). 
The resultant expression takes the form
\begin{align}
\Gamma_{\widetilde{H}^0}(X) &\ni
-2\frac{g'^2_1}{T}
\Big[
	\big(Q_{H_d}^2v_d^2(X)+Q_{H_u}^2v_u^2(X)\big)
	\mathcal{F}_{\widetilde{Z}'\widetilde{H}^0} \non\\
&\hspace{2cm}	
	-Q_{H_d}Q_{H_u}v^2(X)\sin2\beta(X)
	|\mu_{\rm eff}(X)||M'_1|\cos(\delta_\lambda+\delta_{M'_1})
	\mathcal{R}_{\widetilde{Z}'\widetilde{H}^0}
\Big]\mu_{\widetilde{H}^0},
\end{align}
where $\mu_{\widetilde{H}^0}$ denotes the chemical potential of the neutral Higgsino.
The explicit forms of $\mathcal{F}_{\widetilde{Z}'\widetilde{H}^0}$
and $\mathcal{R}_{\widetilde{Z}'\widetilde{H}^0}$ are presented in Appendix \ref{app:Tfunc}.

In what follows, we calculate the BAU by solving the quantum diffusion equations. 
To do so, we consider the number densities of the third generations of the quarks/squarks and 
Higgses/Higgsinos~\cite{Cohen:1994ss,Huet:1995sh}
\begin{align}
Q(X) &= n_{t_L}+n_{\tilde{t}_L}+n_{b_L}+n_{\tilde{b}_L}, \\
T(X) &= n_{t_R}+n_{\tilde{t}_R}, \\
B(X) &= n_{b_R}+n_{\tilde{b}_R}, \\
H(X) &= n_{H_u^+}+n_{H_u^0}+n_{H_d^+}+n_{H_d^0}+n_{\widetilde{H}^+}+n_{\widetilde{H}^0},
\end{align}
where the supergauge equilibrium is assumed. 
For later use, we define the $k$ factors as
\begin{align}
k_{b,f}\left(\frac{m}{T}\right) = 
\frac{6n_{b,f}}{T^2\mu}=\frac{6g}{T^2\mu}\int\frac{d^3\boldsymbol{k}}{(2\pi)^3}
\left[\frac{1}{e^{(\omega-\mu)/T}\mp 1}-\frac{1}{e^{(\omega+\mu)/T}\mp 1}\right],
\end{align}
where $g$ counts the degrees of freedom, and $\mu$ denotes a chemical potential
and $\omega=\sqrt{|\boldsymbol{k}|^2+m^2}$.
Since the wall thickness is much smaller than the wall radius, we can ignore the 
curvature of the bubbles. We thus concentrate on a direction (denoted as $z$ in the plasma frame) 
in which the bubble walls move.
Assuming the particle changing rates via the Yukawa interactions and the strong sphaleron 
(denoted as $\Gamma_Y$ and $\Gamma_{ss}$ respectively) are greater
than those via the Higgs bubble walls (denoted as $\Gamma_M^\pm$ and $\Gamma_h$ which are given below), 
the coupled diffusion equations with respect to $Q(z)$, $T(z)$ and $H(z)$ are reduced to a single diffusion equation
\begin{eqnarray}
v_wH'(\bar{z})-\bar{D}H''(\bar{z})+\bar{\Gamma}(\bar{z})H(\bar{z})-\bar{S}(\bar{z})
+\mathcal{O}\left(\frac{1}{\Gamma_{\rm ss}}, \frac{1}{\Gamma_Y}\right)=0,\label{QDE_H}
\end{eqnarray}
where we transform the plasma frame to the wall rest frame ($z\to \bar{z}=z+v_wt$) with $v_w$ being the
wall velocity. In this frame, $\bar{z}<0$ corresponds to the symmetric phase and $\bar{z}>0$ does to the broken phase.
$\bar{D}$, $\bar{\Gamma}$ and $\bar{S}$ are, respectively, given by
\begin{align}
\bar{D} &= \frac{bD_q+aD_h}{a+b}, \\
\bar{\Gamma} &= \frac{1}{a+b}
\left[
	\frac{a}{k_H}(\Gamma_M^-+\Gamma_h)-(9k_Q-9k_T+3K_B)\Gamma_M^+
\right], \\
\bar{S} &= \frac{a}{a+b}(S_{\tilde{t}}+S_{\tilde{H}}),
\end{align}
where $a = k_H(9k_Q+9k_T+k_B)$, $b = 9k_Qk_T+k_Qk_B+4k_Tk_B$ and 
$D_{q,h}$ are the diffusion constants of the quarks/squarks and Higgses/Higgsinos.
$\Gamma_M^\pm$ and $\Gamma_h$ are defined by
\begin{align}
\Gamma_M^\pm = \frac{6}{T^2}(\Gamma_t^\pm+\Gamma_{\tilde{t}}^\pm), \quad
\Gamma_h = \frac{6}{T^2}(\Gamma_{\widetilde{H}^\pm}+\Gamma_{\widetilde{H}^0}),
\end{align}
where $\Gamma_{t}^\pm$, $\Gamma_{\tilde{t}}^\pm$ and $\Gamma_{\widetilde{H}^\pm}$
are the chirality-flipping rates appearing in the diffusion equations for 
the top, stop and charged Higgsino, respectively.
The calculation of $\Gamma_{t}^\pm$ is somewhat lengthy since the so-called {\it hole} modes
should be taken into account. 
For the details, refer to Ref.~\cite{Lee:2004we}.

Provided that $\bar{\Gamma}(\bar{z})$ is nonzero and constant for $\bar{z}>0$, 
the solution for $H(\bar{z})$ in the symmetric phase is found to be
\begin{eqnarray}
H(\bar{z}) = \mathcal{A}e^{v_w\bar{z}/\bar{D}},\quad
\mathcal{A}=\frac{1}{\bar{D}\lambda_+}\int_0^\infty dz'~\bar{S}(z')e^{-\lambda_+z'},\quad
\lambda_+ = \frac{v_w+\sqrt{v_w^2+4\bar{D}\bar{\Gamma}}}{2\bar{D}}.\label{A}
\end{eqnarray}

The total left-hand number density ($n_L(\bar{z})$) accumulated in the symmetric phase
can be the source for baryogenesis.
The left-handed number densities of the first and second generation fermions are
related to that of the third generation fermions through the strong sphaleron process.
Thus, $n_L(\bar{z})$ is expressed as~\cite{Cohen:1994ss,Huet:1995sh}
\begin{align}
n_L(\bar{z}) &= 5Q(\bar{z})+4T(\bar{z})
=-\left[
	r_1+\frac{r_2v_w^2}{\Gamma_{ss}\bar{D}}
\left(
	1-\frac{D_q}{\bar{D}}
\right)
\right]H(\bar{z}),
\end{align}
where
\begin{align}
r_1 = \frac{9k_Qk_T-5k_Qk_B-8k_Tk_B}{a},\quad
r_2 = \frac{k_Hk_B^2(5k_Q+4k_T)(k_Q+2k_T)}{a^2}.\label{r1r2}
\end{align}

With $n_L(\bar{z})$, the baryon number density can be estimated by the following diffusion equation
\begin{eqnarray}
D_qn''_B(\bar{z})-v_wn'_B(\bar{z})-\theta(-\bar{z})\mathcal{R}n_B(\bar{z})
	=\theta(-\bar{z})\frac{N_g}{2}\Gamma_B^{(s)} n_L(\bar{z}),
\end{eqnarray}
where the relaxation term is given by
$\mathcal{R}= \Gamma_B^{(s)}[9/\{4(1+n_{\rm sq}/6)\}+3/2]$ with $n_{\rm sq}$ 
for the number of light squark flavors~\cite{Cline:2000nw}.
$N_g$ is the number of the fermion generation and $\Gamma_B^{(s)}$ is the baryon number changing
rate in the symmetric phase. 
After imposing the boundary conditions, $n_B(\bar{z}\to -\infty)\to 0$ and $n_B'(\bar{z}>0)=0$, 
one arrives at
\begin{eqnarray}
n_B(\bar{z}>0) = \frac{-N_g\Gamma_B^{(s)}}{2\sqrt{v_w^2+4\mathcal{R}D_q}}
\int_{-\infty}^0dz'~n_L(z')e^{-\kappa_-z'},\label{nB}
\end{eqnarray}
where
\begin{eqnarray}
\kappa_- = \frac{v_w-\sqrt{v_w^2+4\mathcal{R}D_q}}{2D_q}.
\end{eqnarray}
Note that if all squarks are decoupled, one may get $k_Q=6$ and $k_T=k_B=3$
in the massless quark approximation, 
leading to $r_1=0$~\cite{Giudice:1993bb,Huet:1995sh}.
Note that this cancellation would be incomplete if the thermal masses of the quarks were taken into account.
We will quantify Eq.~(\ref{nB}) in the next section.

\section{Numerical analysis}\label{sec:numerics}
There are six free parameters in the tree-level Higgs sector: 
\begin{align}
|\lambda|,\quad |A_\lambda|,\quad \tan\beta,\quad v_S; \quad Q_{H_d}, \quad Q_{H_u}.
\end{align}
We replace $(|\lambda|, |A_\lambda|, v_S)$ with ($m_{H_1}, m_{H^\pm}, m_{Z'}$) respectively
and take the following as the input parameters:
\begin{align}
m_{H_1}(=126~{\rm GeV}),\quad m_{H^\pm}, \quad \sqrt{\frac{|Q_{H_d}|}{|Q_{H_u}|}}, \quad
m_{Z'}; \quad Q_{H_d}, \quad Q_{H_u}.
\end{align}
In our analysis, as a sample point we take $Q_{H_d}=Q_{H_u}=-1/2$, which gives $\tan\beta=1$.
So the $m_{H^\pm}$ and $m_{Z'}$ are the only varying parameters.
The dependence of $Q_{H_d}$ and $Q_{H_u}$ on the results will be discussed below.
We fix the other parameters as
\begin{align}
&m_{\tilde{q}}=m_{\tilde{t}_R}=m_{\tilde{b}_R}=1500~{\rm GeV}, \quad
A_t=A_b=m_{\tilde{q}}+|\mu_{\rm eff}|/\tan\beta, \\
& |M_1|=|M_2|/2=100~{\rm GeV},\quad |M'_1|=|\mu_{\rm eff}|; \\
& \delta_{M_1}=\delta_{M_2}=\delta_{\lambda}=0,\quad \delta_{M_1'}=\pi/2.
\end{align}

\begin{figure}[t]
\center
\includegraphics[width=7.5cm]{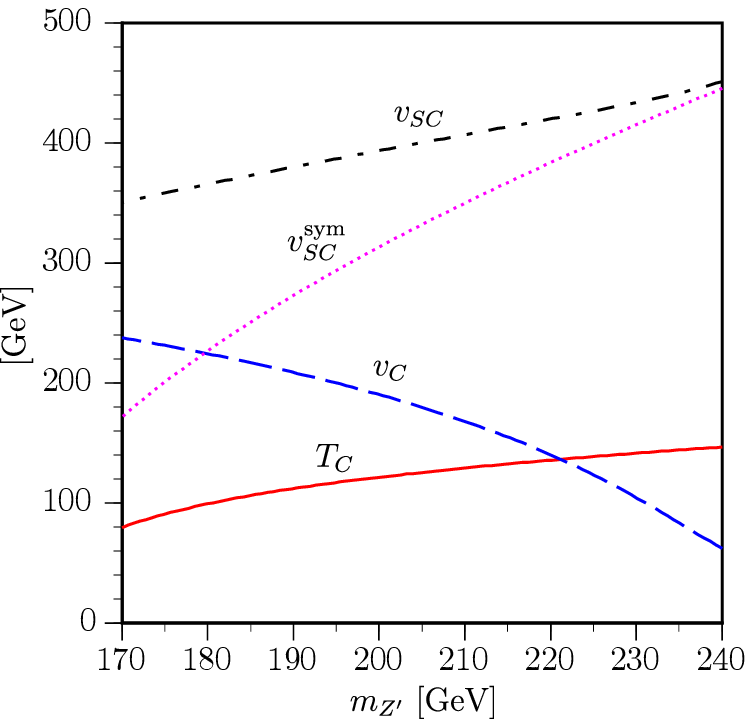}
\hspace{1cm}
\includegraphics[width=7.6cm]{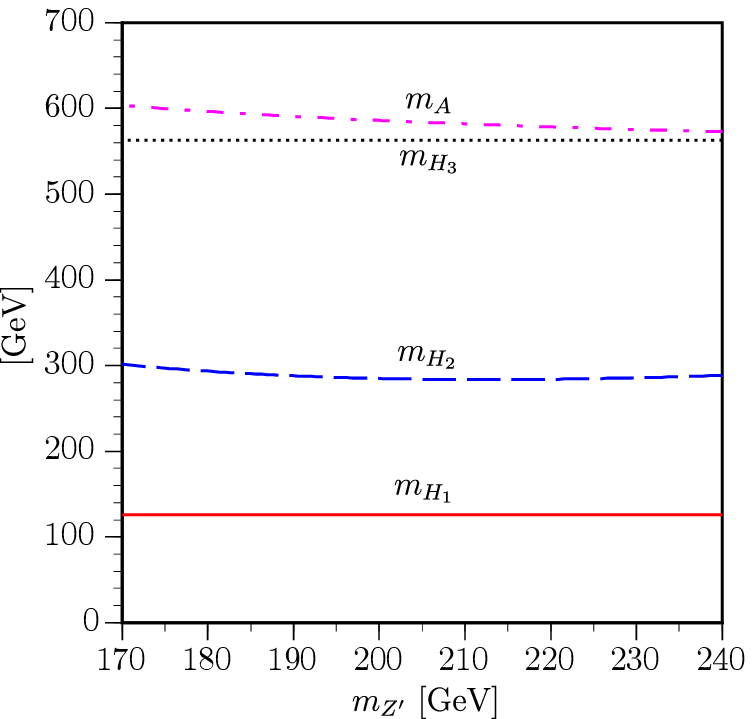}\\[1cm]
\includegraphics[width=7cm]{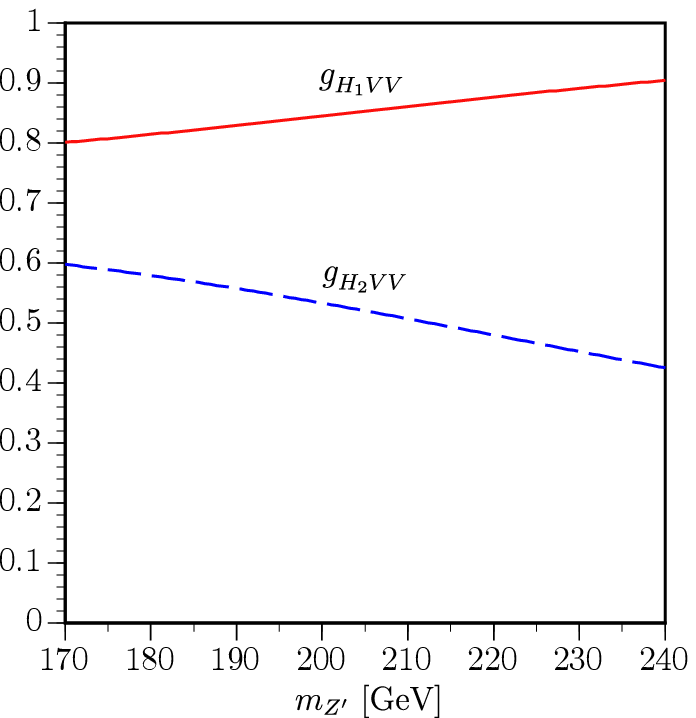}
\hspace{1cm}
\includegraphics[width=7.5cm]{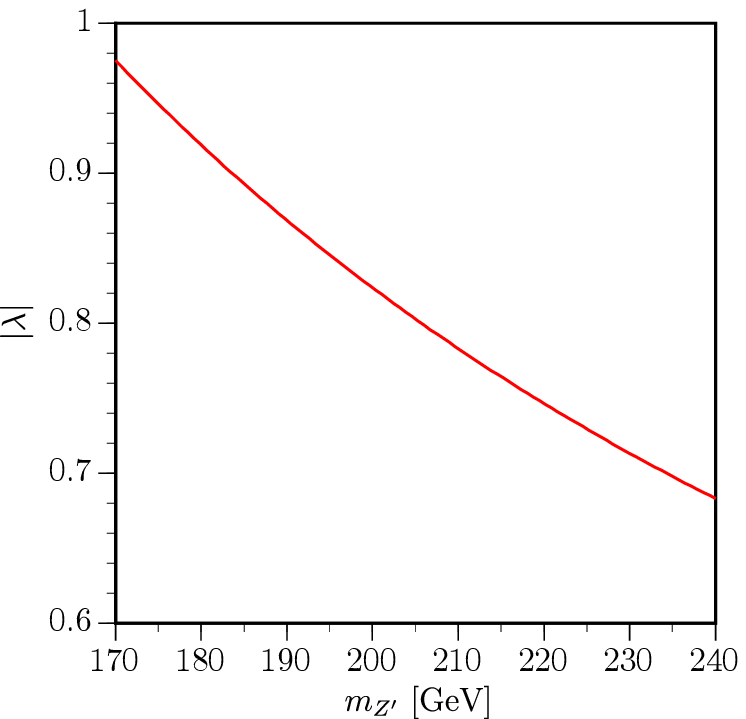}
\caption{We take $m_{H^\pm}=550$ GeV. 
The Higgs VEVs at $T_C$ (Upper left panel), 
the neutral Higgs boson masses (Upper right panel), 
the lightest and second lightest Higgs boson couplings with the gauge bosons (Lower left panel), 
$|\lambda|$ (Lower right panel) are shown as a function of $m_{Z'}$.}
\label{fig:bm1_mzp}
\end{figure}
In the upper left panel of Fig.~\ref{fig:bm1_mzp}, 
the Higgs VEVs at $T_C$ are displayed as a function of $m_{Z'}$,
and $m_{H^\pm}=550$ GeV is taken.
It is found that as $m_{Z'}$ decreases $v_C/T_C$ increases. 
This is because the effect of the doublet-singlet Higgs mixing gets enhanced as $v_S$ decreases.
In this specific example, $v_C/T_C$ can reach about 3 at $m_{Z'}=170$ GeV. 
The reduction of $T_C$ may be the prominent feature of the first-order EWPT 
driven by the doublet-singlet Higgs mixing.
In such a case, the value of the singlet Higgs VEV significantly changes during the EWPT,
leading to large $|v_{SC}-v_{SC}^{\rm sym}|$.
Conversely, in the large $v_S$ limit, the effect of the singlet Higgs field is suppressed and hence
$v_C/T_C$ is weakened.

In the upper right panel of Fig.~\ref{fig:bm1_mzp}, we show the four neutral Higgs boson masses.
We find that $m_{H_2}\simeq 300$ GeV and the other two heavy neutral Higgs boson masses are mostly
controlled by the scale of $R_\lambda v_S$ which is fixed by the charged Higgs boson mass, $m_{H^\pm}=550$ GeV.

In the lower left panel of Fig.~\ref{fig:bm1_mzp}, $g_{H_1VV}$ and $g_{H_2VV}$ are shown, where $g_{H_iVV}$ is
defined by
\begin{align}
g_{H_iVV}^{} = O_{1i}\cos\beta+O_{2i}\sin\beta,
\end{align}
with $O$ being the orthogonal matrix that diagonalizes the neutral Higgs boson mass matrix.
In the SM-like limit, $g_{H_1VV}^{}\to 1$. 
Since $H_1$ and $H_2$ are the mixture of the doublet and singlet Higgs bosons,
$g_{H_1VV}^{}$ can deviate from unity.
In the chosen parameter space, the main component of $H_1$ is the doublet Higgs boson 
while that of $H_2$ is the singlet Higgs boson.
As mentioned above, the smaller $m_{Z'}$, the more doublet-singlet Higgs mixing gets enhanced, 
leading to the stronger first-order EWPT.

In our analysis, $m_{H_1}$ is fixed as 126 GeV by tuning $|\lambda|$.
Note, however,  that the adjusted $|\lambda|$ can vary as $v_S$ changes.
If the doublet-singlet Higgs mixing is large, $m_{H_1}$ tends to decrease, which can be seen 
from the approximate mass formula Eq.~(\ref{mH12_tree}).
In order to set $m_{H_1}=126$ GeV, such a deficit in $m_{H_1}$ should be compensated by the increment
of $|\lambda|$, which explains the behavior of $|\lambda|$ as a function of $m_{Z'}$ 
presented in the lower right panel of Fig.~\ref{fig:bm1_mzp}.
For $m_{Z'}=170$ GeV, $|\lambda|\simeq 0.97$ is needed, 
and a lighter $Z'$ boson would require $|\lambda|>1$
\footnote{In the NMSSM and nMSSM, $|\lambda|<(0.7-0.8)$ should be satisfied 
to avoid a Landau pole below 
a grand unification scale ($\sim10^{16}$ GeV)~\cite{Menon:2004wv,Miller:2003ay}.
If we impose the same bound, the region where the strong first-order EWPT is possible 
would be mostly ruled out. However, the upper bound of $|\lambda|$ may change depending
on a particle content of a full theory that we do not specify.  
In this analysis, we do not impose specific perturbativity bound on $|\lambda|$ and vary it up to 1.  
}.
Note that this statement is based on the assumption of $Q_{H_d}=Q_{H_u}=-1/2$.
The discussion of the different choices of $Q_{H_{d,u}}$ will be given below.

Summarizing our findings in Fig.~\ref{fig:bm1_mzp},
to realize the strong first-order EWPT in the UMSSM, 
the non-MSSM-like limit is needed, which leads to the light $Z'$ boson\footnote{If the $Z'$ boson gets its mass from the additional singlet Higgs bosons such as the one in the sMSSM, 
the relationship between the strength of $v_C/T_C$ and $m_{Z'}$ is not necessarily correlated
 (see e.g. \cite{EWBG_sMSSM,Chiang:2009fs}).}. 
It is found that $m_{Z'}<220$ GeV if $\zeta_{\rm sph}=1$.
The more precise upper bound of $m_{Z'}$ requires the knowledge of
$\zeta_{\rm sph}$ which will be evaluated below.
Since the experimental lower bounds on $m_{Z'}$ in various $Z'$ models
are typically multi-TeV, the EWBG may not be successful except the so-called leptophobic $Z'$ scenario
in which the $Z'$ boson does not or much weakly couple to the leptons, and thus the collider bounds on $m_{Z'}$ 
may be significantly relaxed.
We will discuss the possible experimental bounds on the leptophobic $Z'$ boson 
in \ref{subsec:exp}.

\begin{figure}[t]
\center
\includegraphics[width=7.5cm]{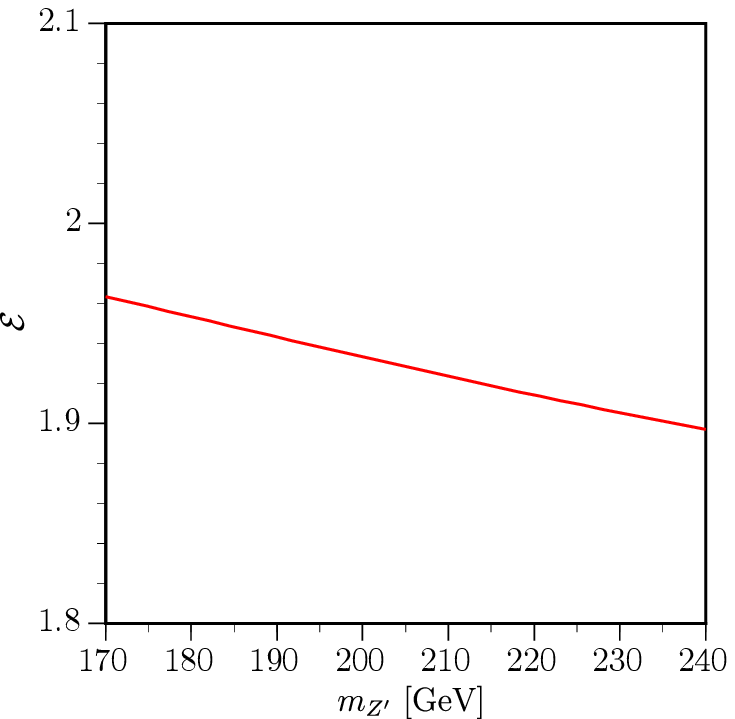}
\hspace{1cm}
\includegraphics[width=7cm]{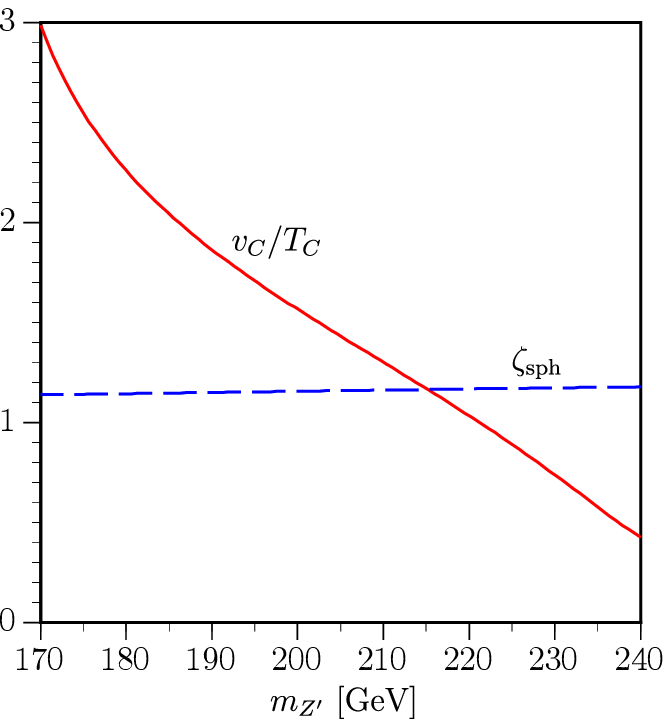}
\caption{(Left panel) $\mathcal{E}$ as a function of $m_{Z'}$. (Right panel) The comparison of 
$v_C/T_C$ with $\zeta_{\rm sph}$. The input parameters are the same as in Fig.~\ref{fig:bm1_mzp}}
\label{fig:rEsph_mzp}
\end{figure}
In the left panel of Fig.~\ref{fig:rEsph_mzp}, $\mathcal{E}$ is plotted as a function of $m_{Z'}$. 
We can see that $\mathcal{E}$ varies from 1.96 to 1.90 in the range $m_{Z'}\in [170, 240]$ GeV.
This behavior may be explained by the fact that the sphaleron energy is sensitive to the
magnitude of $|\lambda|$ as is observed in the SM sphaleron case,
namely, the larger the $|\lambda|$ yields, the larger the sphaleron energy.

The right panel of Fig.~\ref{fig:rEsph_mzp} shows $\zeta_{\rm sph}$ (blue dashed line)
as a function of $m_{Z'}$,
where the only leading correction is retained in Eq.~(\ref{sph_dec}).
It is found that $\zeta_{\rm sph}\in(1.14,1.18)$. 
Here, we also overlay $v_C/T_C$ (red straight line) and find that
$v_C/T_C>\zeta_{\rm sph}$ is satisfied for $m_{Z'}< 215$ GeV.
As noted in Sec.~\ref{sec:sphaleron}, the sphaleron decoupling condition evaluated with $\mathcal{E}(0)$
should be improved by other effects. If we adopt the MSSM result, 
$\zeta_{\rm sph}=1.4$~\cite{Funakubo:2009eg}, 
we would have $m_{Z'}< 206$ GeV.

\begin{figure}[t]
\center
\includegraphics[width=7.5cm]{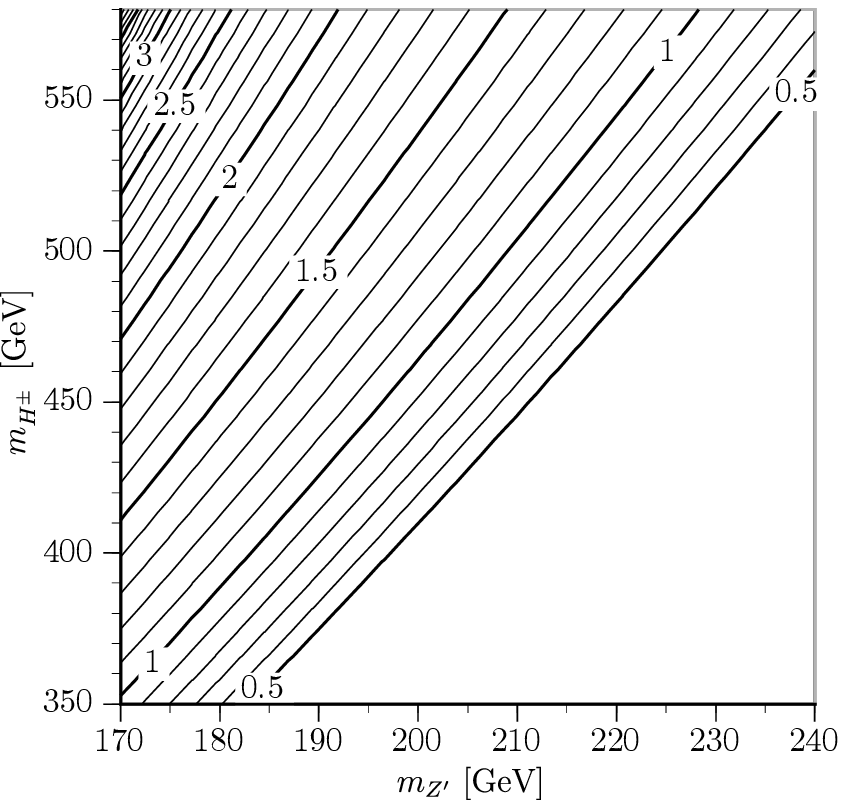}
\hspace{1cm}
\includegraphics[width=7.5cm]{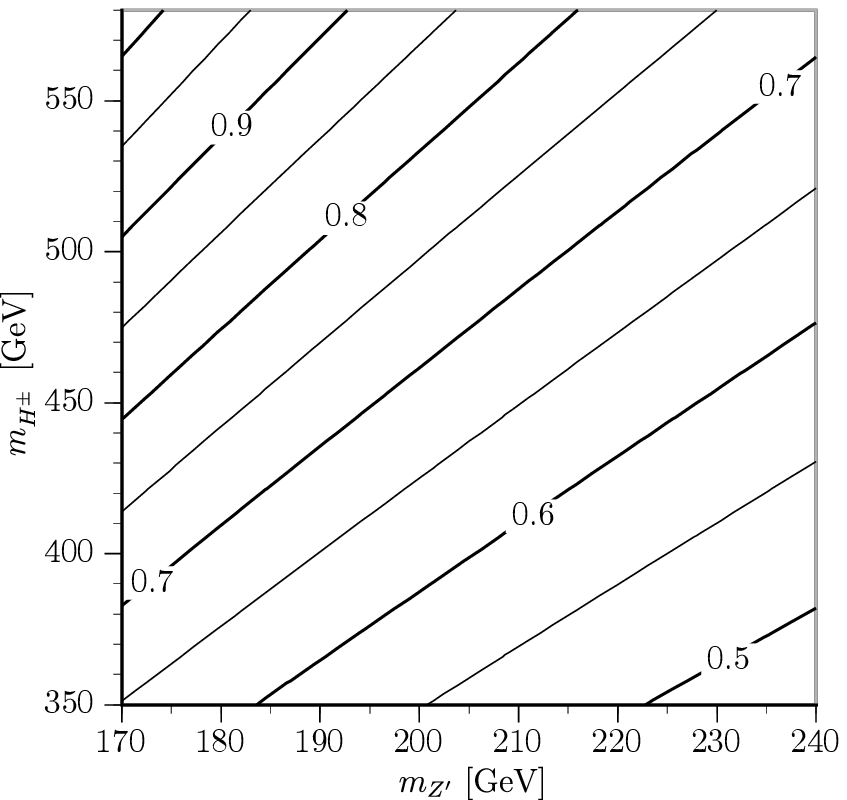}
\caption{$v_C/T_C$ (Left panel) and $|\lambda|$ (Right panel) are plotted in the $m_{H^\pm}$-$m_{Z'}$ plane.}
\label{fig:vcTc_lam_mzp_mch}
\end{figure}
So far, we present the results only in the case of $m_{H^\pm}=550$ GeV. 
Here, we study the dependence of $m_{H^\pm}$ on the strength of the first-order EWPT.
In the left panel of Fig.~\ref{fig:vcTc_lam_mzp_mch}, the contours of $v_C/T_C$ are plotted
in the $m_{H^\pm}$-$m_{Z'}$ plane.
Since $|A_\lambda|$ dictates the magnitude of 
the singlet-doublet Higgs mixing effect, the larger $m_{H^\pm}$ can give the stronger first-order EWPT.
However, we should note that there is a maximal value of $m_{H^\pm}$ with a fixed $m_{Z'}$ (or $v_S$)
from the vacuum metastability as discussed in \ref{subsec:Veff1}.

$|\lambda|$ is plotted in the right panel of Fig.~\ref{fig:vcTc_lam_mzp_mch}. 
Similar to the behavior of $|\lambda|$ as a function of $m_{Z'}$, 
the larger $|\lambda|$ is needed to realize $m_{H_1}=126$ GeV as $m_{H^\pm}$ increases.

After scanning $m_{H^\pm}$ and $m_{Z'}$ in wider parameter space,
it is found that $v_C/T_C>\zeta_{\rm sph}$ can be satisfied for $m_{Z'}\lesssim (150-300)$ GeV.

Before going on to the BAU estimates, we discuss the effects of $Q_{H_{d,u}}$ on the strength of the first-order EWPT.
As $Q_{H_{d,u}}$ decrease, $Q_S$ becomes smaller through $Q_S=-(Q_{H_d}+Q_{H_u})$.
For such a smaller $Q_S$, $v_S$ has to be larger for a fixed $m_{Z'}$ through Eq.~(\ref{mZ-mZp}),
and, correspondingly, $R_\lambda$ decreases for a fixed $m_{H^\pm}$ as can be seen from Eq.~(\ref{mch_tree}).
As we have discussed above, the EWPT gets weakened as $v_S$ increases
and/or $R_\lambda$ decreases.
We also note that as the singlet-doublet Higgs mixing effect gets smaller, the value of $|\lambda|$ also becomes smaller to set $m_{H_1}=126$ GeV.
From the above discussion, to realize the strong-first EWPT in the case of the smaller $Q_{H_{d,u}}$, 
the smaller $m_{Z'}$ and/or the larger $m_{H^\pm}$ would be needed,
and it is the other way around for the larger $Q_{H_{d,u}}$.

\begin{figure}[t]
\center
\includegraphics[width=7.5cm]{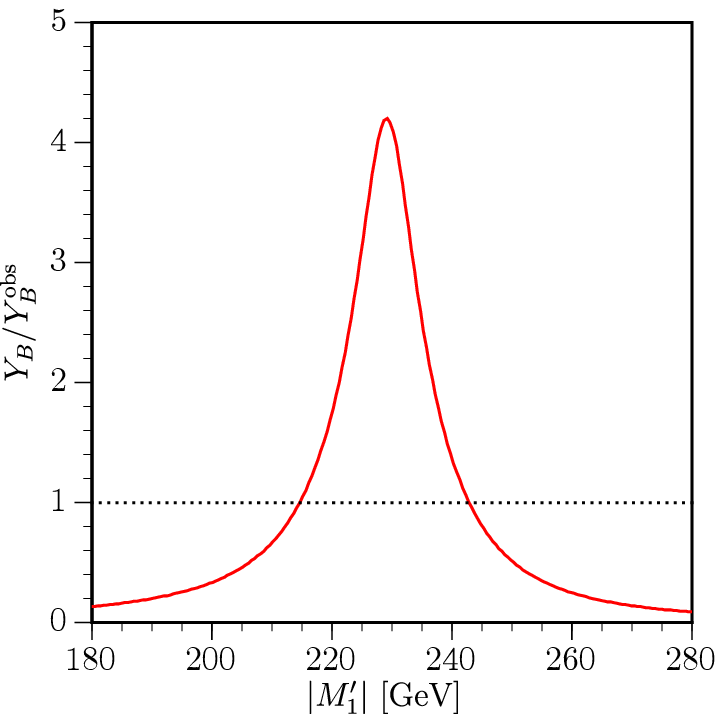} 
\hspace{1cm}
\includegraphics[width=7.5cm]{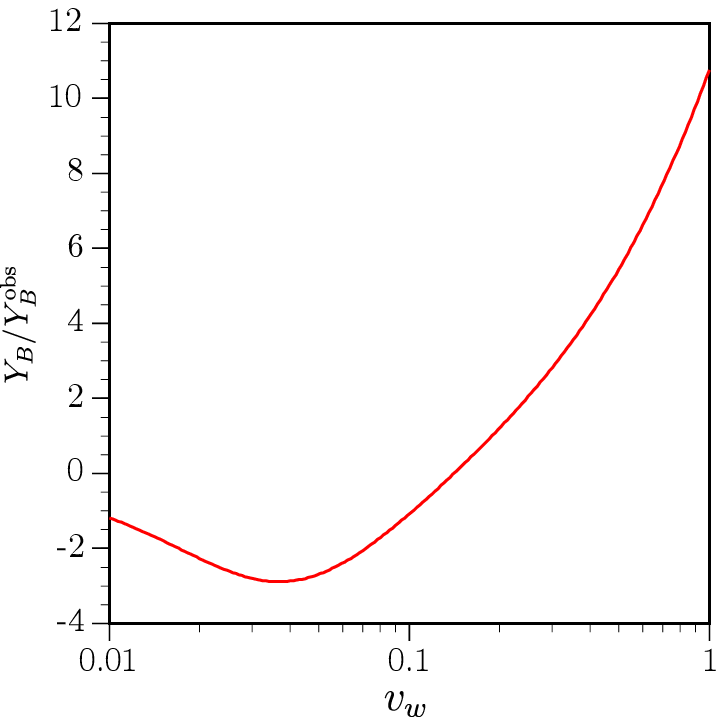}
\caption{(Left panel) $Y_B/Y_B^{\rm obs}$ is plotted as a function of $|M_1'|$. 
The input parameters are the same as in Fig.~\ref{fig:bm1_mzp} but fixed $m_{Z'}=200$ GeV.
We take $v_w=0.4$ as the bubble wall velocity and other relevant parameters are given in the text.
(Right panel) The $v_w$ dependence on $Y_B/Y_B^{\rm obs}$ is shown.}
\label{fig:YB_M1p_vw}
\end{figure}
Now let us estimate the BAU. We consider the case presented in Fig.~\ref{fig:bm1_mzp}
choosing $m_{Z'}=200$ GeV. In this case, the EWPT takes place as
\begin{align}
(v_C^{\rm sym}, v_{SC}^{\rm sym}) = (0~{\rm GeV}, 313.4~{\rm GeV}) \to
(v_C, v_{SC}) = (190.3~{\rm GeV}, 393.6~{\rm GeV})
\end{align}
at $T_C=121.5$ GeV, giving rise to $v_C/T_C>\zeta_{\rm sph}$.

In order to proceed to the BAU estimates,
we assume the following kink-type bubble wall profiles for the sake of simplicity,
\begin{align}
v(\bar{z}) 
&= \frac{v_C}{2}\left[1-\tanh\left\{\alpha\left(1-\frac{2\bar{z}}{L_w}\right)\right\}\right],\\
\beta(\bar{z}) &= \beta^{({\rm br})}(T_C)-\frac{\Delta\beta}{2}
	\left[1+\tanh\left\{\alpha\left(1-\frac{2\bar{z}}{L_w}\right)\right\}\right],\\
v_S(\bar{z}) &= v_{SC}-\frac{1}{2}\big(v_{SC}-v_{SC}^{{\rm sym}}\big)
	\left[1+\tanh\left\{\alpha\left(1-\frac{2\bar{z}}{L_w^{(S)}}\right)\right\}\right],
\end{align}
where we take the following as reference values:
\begin{align}
\alpha=3/2,\quad \Delta \beta=0.01, \quad L_w=L_w^{(S)}=\frac{5}{T}.
\end{align}
The other relevant parameters are fixed as
\begin{align}
\Gamma_{\widetilde{H}^0} &=0.025T, 
\quad \Gamma_{\widetilde{Z}'}=0.003T,\quad
D_q=\frac{6}{T},\quad D_h=\frac{110}{T},\\
\Gamma_B^{(s)} &= 6\kappa\alpha_W^4T,\quad \Gamma_{ss} = 16\kappa'\alpha_s^4T,
\end{align}
with $\alpha_W=g_2^2/4\pi$, $\alpha_s=g_s^2/4\pi$, $\kappa=20$ and $\kappa'=1$.

In the left panel of Fig.~\ref{fig:YB_M1p_vw}, we plot $Y_B/Y_B^{\rm obs}$ as a function of the $Z'$-ino mass, where
$Y_B^{\rm obs}=n_B/s=8.57\times 10^{-11}$~\cite{Ade:2013zuv},
and $v_w=0.4$~\cite{Moore:1995si} is taken for the bubble wall velocity.
It is found that the sufficient BAU can be generated if $|M_1'|\simeq |\mu_{\rm eff}(T_C)|\simeq 229.2$ GeV.
Such a resonant enhancement is well studied in the MSSM known as \lq\lq Wino- and Bino-driven EWBG\rq\rq,
where the Wino or Bino mass is almost degenerate with the Higgsino mass
\footnote{The \lq\lq Singlino-driven EWBG\rq\rq is considered in the NMSSM~\cite{Cheung:2012pg}. 
However, such a scenario relies on the off-resonance effect rather than resonance one
since the neutral Higgsino and singlino masses cannot be degenerate due to the successful condition for the strong first-order EWPT (called type-B EWPT~\cite{Funakubo:2005pu}).}.
Note that in the \lq\lq $Z'$-ino-driven EWBG\rq\rq investigated here, 
the mass degeneracy between the Higgsino and $Z'$-ino
occurs dynamically during the EWPT rather than from the onset.

As studied in Refs.~\cite{vw_LSS}, 
if the lighter stop is light enough, $v_w$ tends to decrease, for instance, $v_w\simeq 0.05$ in the MSSM.
In the right panel of Fig.~\ref{fig:YB_M1p_vw}, the wall velocity dependence on the BAU is displayed.
In the chosen parameter set, the magnitude of the BAU is greater than one except $0.1\lesssim v_w\lesssim 0.2$.

It is important to note that the BAU linearly depends on $\Delta\beta$.
In the MSSM, $\Delta\beta$ can vary in the range $\mathcal{O}(10^{-2}-10^{-3})$
depending on the $CP$-odd Higgs boson mass and $\tan\beta$~\cite{Funakubo:2009eg,Moreno:1998bq}.
In the UMSSM, on the other hand, $\Delta\beta$ is not known, and so the variation 
of $\Delta\beta$ is regarded as the theoretical uncertainty in the current investigation.
Our reference value of $\Delta\beta$ inferred by the MSSM analysis could be optimistic. 
If we take $\Delta\beta=0.001$, the generated BAU is reduced by a factor of 10,
which is somewhat below the observed value.
In this case, the deficit has to be compensated by other effects, 
for example, increment of the $U(1)'$ charges,
and it seems possible even after taking account of the experimental constraints on them,
which will be discussed below.
In any case, the BAU generated by the $Z'$-ino effect can fall in the right ballpark
within theoretical uncertainties. 
The estimate of the more accurate BAU requires full knowledge of the Higgs bubble wall profiles.
We defer the detailed analysis to future work.

%
%
\subsection{Experimental constraints}\label{subsec:exp}
As discussed above, the strong-first order EWPT can be realized in the light $Z'$ boson region.
Such a $Z'$ boson is constrained by collider searches and electroweak precision tests. 
Even though the $Z'$ boson is leptophobic, the dijet-mass searches 
may give the useful constraints (For a recent study, see e.g. \cite{Dobrescu:2013cmh}).
At the Tevatron and LHC, however, the dijet-mass searches are limited to $M_{jj}>200$ GeV,
and thus there is not enough sensitivity to search for the $Z'$ boson with a mass less than around 200 GeV. In this mass region, the stringent collider bounds may come from the UA2 experiments~\cite{Alitti:1993pn}.
In Ref.~\cite{Buckley:2011vc}, using the UA2 data it is found that $g_{u_iu_iZ'}^{}\lesssim0.2-0.5$ for $130~{\rm GeV}\lesssim m_{Z'}\lesssim 300$ GeV, where $g_{u_iu_iZ'}\bar{u}_i\gamma^\mu u_iZ'_\mu$ with $i$ being chiralities. 
In our notation, $g_{f_if_iZ'}=g'_1Q_{f_i}$.
Since $U(1)'$ charge assignments of the matter sector are highly model dependent, 
we do not specify them explicitly. 
As an example, if we take $Q_{q_L^{}}=Q_{q_R^{}}=0.25~(=-Q_{H_{d,u}}/2)$,
we obtain $g_{qqZ'}\simeq 0.11$, which is below the UA2 experimental bounds.

In Refs.~\cite{Umeda:1998nq}, the constraints on leptophobic $Z'$ models 
from the electroweak precision tests are discussed. 
In order to have the viable $Z'$ boson with the mass range discussed in this paper, 
the $Z$-$Z'$ mixing has to be highly suppressed. 
This may be possible if $Q_{H_{d,u}}$ and $\tan\beta$ are chosen in such a way that
the $Z$-$Z'$ mixing is zero. As discussed in \ref{subsec:mH_mZp},
such a choice can be easily made at the tree level, namely, $\tan\beta=\sqrt{|Q_{H_d}|/|Q_{H_u}|}$.
However, since the nonzero $Z$-$Z'$ mixing appears at the loop level, 
the more careful choices of $Q_{H_{d,u}}$ and $\tan\beta$ may be needed.
Nevertheless, we expect that since the deviation from the tree-level relation is loop induced, 
such a modification would not change our results drastically.

In our scenario, the $CP$-violating phase relevant to the EWBG is $\delta_\lambda+\delta_{M'_1}$.
Such a phase can contribute to the electric dipole moments of Thallium, neutron and Mercury etc.
However, the constraints are more or less the same as those in the Bino-driven EWBG~\cite{Li:2008ez},
which are less stringent even in the case of the maximal $CP$ violation.

As discussed above, a significant amount of the doublet-singlet Higgs mixing is needed 
to have the strong first-order EWPT, which implies that $g_{H_1VV}^{}$ inevitably deviates from
its SM value as demonstrated in Fig.~\ref{fig:bm1_mzp}. 
However, our $g_{H_1VV}^{}$ is still within the 2$\sigma$ ranges of the current LHC data~\cite{gHVV_LHCdata}.

\section{Conclusions and Discussion}\label{sec:conclusion}
We have examined the feasibility of the EWBG in the UMSSM in light of the 126 GeV Higgs boson.
It is found that the EWPT is strongly first-order due to the doublet-singlet Higgs mixing.
The degree to which the doublet-singlet Higgs mixing is enhanced depends on the scale of $v_S$.
In the region where the strong first-order EWPT is realized, the $Z'$ boson is necessarily light.
For a typical parameter set, we found $\zeta_{\rm sph}=(1.1-1.2)$, and
$v_C/T_C>\zeta_{\rm sph}$ is satisfied if $m_{Z'}\lesssim 215$ GeV.
In this mass range, the $Z'$ boson has to be leptophobic to be consistent with the collider bounds.
After scanning the relevant parameters, $m_{H^\pm}$ and $m_{Z'}$,
it is found that $v_C/T_C>\zeta_{\rm sph}$ can be satisfied for $m_{Z'}\lesssim (150-300)$ GeV.

We also estimated the BAU based on the CTP formalism, and showed that the $CP$ violation arising from
the interactions between the $Z'$-ino and neutral Higgsino plays a crucial role in generating the BAU.
Similar to the Wino- and Bino-driven EWBG scenarios in the MSSM, the masses of 
the $Z'$-ino and neutral Higgsino should be degenerate to enhance the BAU,
which occurs dynamically during the EWPT in the $Z'$-ino-driven EWBG scenario.

Since our analysis is conducted in the bottom-up approach, 
the $U(1)'$ charges are not fixed by a concrete UV theory.
It should be noted that in the case of the vanishing lepton $U(1)'$ charges, 
the lepton masses cannot be generated through the ordinary Yukawa interaction 
$\epsilon_{ij}f_{AB}^{(e)}\widehat{H}_d^i\widehat{L}_A^j\widehat{E}_B$ provided that $Q_{H_d}\neq0$
which is required for the $Z'$-ino-driven EWBG scenario.
However, if there is an additional Higgs doublet that couples to the leptons through the Yukawa interactions as written above,
the lepton mass generation issue might be circumvented.  
We defer further investigation to future work. 
Nevertheless, we emphasize that whatever the UV theory is, 
the parameter space that is consistent with the successful EWBG
is expected to be similar to the one we investigate in this paper.

The light leptophobic $Z'$ boson can be tested by future low energy experiments~\cite{future_LowExp}.
For a recent study, see e.g. \cite{FCNC_Zp}.

Since the strong first-order EWPT necessarily leads to the deviation of $g_{H_1VV}^{}$
from the SM value, our scenario can be tested by future coupling measurements 
at the high-luminosity LHC~\cite{HL-LHC} and the International Linear Collider~\cite{ILC}.

\appendix
%
%
\section{Thermal masses of the gauge bosons}\label{app:mT}
Here, we give the thermal masses of the gauge bosons.
For simplicity, we do not consider the mixing terms between the $U(1)_Y$ and $U(1)'$ bosons.
The thermal masses of the $SU(2)_L$, $U(1)_Y$ and $U(1)'$ gauge bosons
to leading order in the high-temperature expansion are, respectively, given by
\begin{align}
\Pi_W(T) &= \Pi_W^{(\rm SM)}(T)+\Pi_W^{(\rm 2nd\;Higgs)}(T)
	+\Pi_W^{(\rm Higgsino)}(T)+\Pi_W^{(\rm Wino)}(T) \non \\
	&=\left[\frac{11}{6}+\frac{1}{6}+\frac{1}{6}+\frac{1}{3}\right]g_2^2T^2=\frac{15}{6}g_2^2T^2, \\
\Pi_B(T) &= \Pi_B^{(\rm SM)}(T)+\Pi_W^{(\rm 2nd\;Higgs)}(T)+\Pi_B^{(\rm Higgsino)}(T) \non\\
	&=\left[\frac{11}{6}+\frac{1}{6}+\frac{1}{6}\right]g_1^2T^2=\frac{13}{6}g_1^2T^2, \\
\Pi_{Z'}(T) &= \Pi_{Z'}^{(\rm Higgs)}(T)+\Pi_{Z'}^{(\rm Higgsino)}(T)
	+\Pi_{Z'}^{(\rm SM\mbox{-}fermion)}(T) \non\\
	&=\frac{2g'^2_1}{3}\left(Q_{H_d}^2+Q_{H_u}^2+\frac{Q_S^2}{2}\right)T^2
	+\frac{g'^2_1}{3}\left(Q_{H_d}^2+Q_{H_u}^2+\frac{Q_S^2}{2}\right)T^2 \non\\
&\quad +N_g\frac{g'^2_1}{6}
	\Big[
		N_C(2Q_Q^2+Q_U^2+Q_D^2)+2Q_L^2+Q_E^2
	\Big]T^2,\label{PiZp}
\end{align}
where we denote ``Higgsino" as $\widetilde{H}^\pm$, $\widetilde{H}^0$, $\widetilde{S}$ and
``Wino" as $\widetilde{W}^\pm$, $\widetilde{W}_3$ and ``Higgs" as $\Phi_d$, $\Phi_u$, $S$.
In our case, $N_g=N_C=3$.
Since we do not specify the $U(1)'$ charges ($Q_Q$ for left-handed quarks, $Q_U$ for up-type right-handed quarks,
$Q_D$ for down-type right-handed quarks, $Q_L$ for left-handed leptons, $Q_E$ for down-type right-handed leptons) 
in the fermion sector, we omit the last line in Eq.~(\ref{PiZp}) in our numerical analysis. 
As mentioned in the text, however, 
the strength of the first-order EWPT is mostly determined by the tree-level structure of the Higgs potential,
and so the above omission does not change our results drastically.
In passing, the SM contributions are decomposed into~\cite{Carrington:1991hz}
\begin{align}
\Pi_W^{(\rm SM)}(T) &= \Pi_W^{\rm (gauge+FP)}(T)
	+\Pi_W^{\rm (SM\mbox{-}Higgs)}(T)+\Pi_W^{\rm (SM\mbox{-}fermion)}(T) \non\\
&=\left[
	\frac{2}{3}+\frac{1}{6}+\frac{N_g}{12}(N_C+1)
	\right]g_2^2T^2, \\
\Pi_B^{(\rm SM)}(T) &= \Pi_B^{\rm (gauge+FP)}(T)
	+\Pi_B^{\rm (SM\mbox{-}Higgs)}(T)+\Pi_B^{\rm (SM\mbox{-}fermion)}(T) \non\\
&=\left[
	0+\frac{1}{6}+\frac{N_g}{12}\left(\frac{11}{9}N_C+3\right)
	\right]g_1^2T^2.
\end{align}
%
%
\section{Thermal functions}\label{app:Tfunc}
The thermal function appearing in the $CP$-violating source term is~\cite{CTP_Riotto}
\begin{align}
\mathcal{I}_{ji}^f&= \frac{1}{4\pi^2}\int_0^\infty dk~\frac{k^2}
	{\omega_j\omega_i}
\Big[
	\big(1-2{\rm Re}(n_i)\big)
	I(\omega_j, \Gamma_j,\omega_i,\Gamma_i)
	+\big(1-2{\rm Re}(n_j)\big)I(\omega_i, \Gamma_i,\omega_j,\Gamma_j)\non\\
&\quad\hspace{4cm}
	-2\big({\rm Im}(n_j)+{\rm Im}(n_i)\big)
	G(\omega_j,\Gamma_j,\omega_i, \Gamma_i)
\Big],
\end{align}
where $n_{i}=1/(e^{(\omega_{i}-i\Gamma_{i})/T}+1)$, $\omega_i=\sqrt{k^2+m_i^2}$, 
$k=|\boldsymbol{k}|$ and $I$ and $G$ are defined by
\begin{align}
I(a,b,c,d) 
&= (b+d)
\left[
	\frac{a+c}{\big\{(a+c)^2+(b+d)^2\big\}^2}
	+\frac{a-c}{\big\{(a-c)^2+(b+d)^2\big\}^2}
\right], \\
G(a,b,c,d) 
&= \frac{1}{2}
\left[
	\frac{(a+c)^2-(b+d)^2}{\big\{(a+c)^2+(b+d)^2\big\}^2}
	-\frac{(a-c)^2-(b+d)^2}{\big\{(a-c)^2+(b+d)^2\big\}^2}
\right].
\end{align}
Note that $G$ is maximized at $a=c$, which implies $m_i=m_j$, and 
$\mathcal{I}_{ji}^f$ vanishes if $\Gamma_i=\Gamma_j=0$.

The thermal functions appearing in the $CP$-conserving chirality-flipping rate is 
\begin{align}
\mathcal{F}_{ji}
&= \frac{1}{2\pi^2}\int_0^\infty dk\frac{k^2}{\omega_{j}\omega_{i}}
\Big[
	{\rm Re}(\tilde{n}_{i})
\big\{	
	J(\omega_{j},\Gamma_{j},\omega_{i},\Gamma_{i})	
	+k^2(\alpha_{ji}^+-\alpha_{ji}^-)
\big\} \non\\
&\quad\hspace{3.5cm}
	+{\rm Im}(\tilde{n}_{i})
\big\{	
	K(\omega_{j},\Gamma_{j},\omega_{i},\Gamma_{i})	
	-k^2(\beta_{ji}^++\beta_{ji}^-)
\big\}, \\
\mathcal{R}_{ji}
&= \frac{-1}{2\pi^2}\int_0^\infty dk\frac{k^2}{\omega_{j}\omega_{i}}
\Big[
	{\rm Re}(\tilde{n}_{i})(\alpha_{ji}^+-\alpha_{ji}^-)
	-{\rm Im}(\tilde{n}_{i})(\beta_{ji}^++\beta_{ji}^-)
\Big],
\end{align}
where $\tilde{n}_i  = n_i(1-n_i)$ and 
\begin{align}
J(\omega_{j},\Gamma_{j},\omega_{i},\Gamma_{i})
&=
	(\omega_{j}\omega_{i}+\Gamma_{j}\Gamma_{i})
	\alpha_{ji}^-
	+(\omega_{j}\omega_{i}-\Gamma_{j}\Gamma_{i})
	\alpha_{ji}^+
	\non\\
&\quad-(\omega_{j}\Gamma_{i}-\Gamma_{j}\omega_{i})
	\beta_{ji}^-
	+(\omega_{j}\Gamma_{i}+\Gamma_{j}\omega_{i})
	\beta_{ji}^+,\\
K(\omega_{j},\Gamma_{j},\omega_{i},\Gamma_{i})
&= (\omega_{j}\omega_{i}+\Gamma_{j}\Gamma_{i})
	\beta_{ji}^-
	-(\omega_{j}\omega_{i}-\Gamma_{j}\Gamma_{i})
	\beta_{ji}^+
	\non\\
&\quad+(\omega_{j}\Gamma_{i}-\Gamma_{j}\omega_{i})
	\alpha_{ji}^-
	+(\omega_{j}\Gamma_{i}+\Gamma_{j}\omega_{i})
	\alpha_{ji}^+,
\end{align}
with
\begin{align}	
\alpha_{ji}^\pm 
= \frac{\Gamma_{j}+\Gamma_{i}}
	{(\omega_{j}\pm \omega_{i})^2
	+(\Gamma_{j}+\Gamma_{i})^2}, \quad
\beta_{ji}^\pm 
= -\frac{\omega_{j}\pm\omega_{i}}
	{(\omega_{j}\pm \omega_{i})^2
	+(\Gamma_{j}+\Gamma_{i})^2}.	
\end{align}

\begin{acknowledgments}
The author thanks Gi-Chol Cho for useful discussions. 
\end{acknowledgments}



\begin{thebibliography}{99}
\bibitem{Beringer:1900zz} 
  J.~Beringer {\it et al.}  [Particle Data Group Collaboration],
  Phys.\ Rev.\ D {\bf 86}, 010001 (2012).

\bibitem{Sakharov:1967dj}
  A.~D.~Sakharov,
  Pisma Zh.\ Eksp.\ Teor.\ Fiz.\  {\bf 5}, 32 (1967)
  [JETP Lett.\  {\bf 5}, 24 (1967\ SOPUA,34,392-393.1991\ UFNAA,161,61-64.1991)].

\bibitem{ewbg}
  V.~A.~Kuzmin, V.~A.~Rubakov and M.~E.~Shaposhnikov,
  Phys.\ Lett.\ B {\bf 155} (1985) 36.
For reviews on electroweak baryogenesis, see
A.~G.~Cohen, D.~B.~Kaplan and A.~E.~Nelson,
Ann.\ Rev.\ Nucl.\ Part.\ Sci.\  {\bf 43} (1993) 27;
%
M.~Quiros,
Helv.\ Phys.\ Acta {\bf 67} (1994) 451;
%
V.~A.~Rubakov and M.~E.~Shaposhnikov,
Usp.\ Fiz.\ Nauk {\bf 166} (1996) 493;
%
K.~Funakubo,
Prog.\ Theor.\ Phys.\  {\bf 96} (1996) 475;
%
M.~Trodden,
Rev.\ Mod.\ Phys.\  {\bf 71} (1999) 1463;
%
W.~Bernreuther,
Lect.\ Notes Phys.\  {\bf 591} (2002) 237;
%
  J.~M.~Cline,
  [arXiv:hep-ph/0609145];
%
  D.~E.~Morrissey and M.~J.~Ramsey-Musolf,
  New J.\ Phys.\  {\bf 14}, 125003 (2012);~
%
  T.~Konstandin,
  arXiv:1302.6713 [hep-ph].

\bibitem{ewbg_sm_cp}
  M.~B.~Gavela, P.~Hernandez, J.~Orloff and O.~Pene,
  Mod.\ Phys.\ Lett.\  A {\bf 9} (1994) 795;~
%
  M.~B.~Gavela, P.~Hernandez, J.~Orloff, O.~Pene and C.~Quimbay,
  Nucl.\ Phys.\  B {\bf 430} (1994) 382;~
%
  P.~Huet and E.~Sather,
  Phys.\ Rev.\  D {\bf 51} (1995) 379;~
%
  T.~Konstandin, T.~Prokopec and M.~G.~Schmidt,
  Nucl.\ Phys.\  B {\bf 679} (2004) 246.

\bibitem{sm_ewpt}
  K.~Kajantie, M.~Laine, K.~Rummukainen and M.~E.~Shaposhnikov,
  Phys.\ Rev.\ Lett.\  {\bf 77}, 2887 (1996);~
%
  K.~Rummukainen, M.~Tsypin, K.~Kajantie, M.~Laine and M.~E.~Shaposhnikov,
  Nucl.\ Phys.\  B {\bf 532}, 283 (1998);~
%
  F.~Csikor, Z.~Fodor and J.~Heitger,
  Phys.\ Rev.\ Lett.\  {\bf 82}, 21 (1999);~
%
  Y.~Aoki, F.~Csikor, Z.~Fodor and A.~Ukawa,
  Phys.\ Rev.\  D {\bf 60}, 013001 (1999).

\bibitem{MSSM-EWBG_LHCtension}  
  T.~Cohen, D.~E.~Morrissey and A.~Pierce,
  Phys.\ Rev.\ D {\bf 86}, 013009 (2012);~
%
  D.~Curtin, P.~Jaiswal and P.~Meade,
  JHEP {\bf 1208}, 005 (2012);~
%
  M.~Carena, G.~Nardini, M.~Quiros and C.~E.~M.~Wagner,
  JHEP {\bf 1302}, 001 (2013);~
 %
  K.~Krizka, A.~Kumar and D.~E.~Morrissey,
  arXiv:1212.4856 [hep-ph].
%

\bibitem{EWBG_NMSSM}
  M.~Pietroni,
  Nucl.\ Phys.\ B {\bf 402}, 27 (1993);~
%
  S.~J.~Huber and M.~G.~Schmidt,
  Nucl.\ Phys.\ B {\bf 606}, 183 (2001);~
%
  J.~Kozaczuk, S.~Profumo and C.~L.~Wainwright,
  arXiv:1302.4781 [hep-ph].

\bibitem{Funakubo:2005pu} 
  K.~Funakubo, S.~Tao and F.~Toyoda,
  Prog.\ Theor.\ Phys.\  {\bf 114}, 369 (2005).

\bibitem{Carena:2011jy} 
  M.~Carena, N.~R.~Shah and C.~E.~M.~Wagner,
  Phys.\ Rev.\ D {\bf 85}, 036003 (2012).
  
\bibitem{Menon:2004wv}
  A.~Menon, D.~E.~Morrissey and C.~E.~M.~Wagner,
  Phys.\ Rev.\  D {\bf 70} (2004) 035005.
%
\bibitem{Huber:2006wf}
  S.~J.~Huber, T.~Konstandin, T.~Prokopec and M.~G.~Schmidt,
  Nucl.\ Phys.\  B {\bf 757} (2006) 172.
  
  \bibitem{EWBG_UMSSM}
  S.~W.~Ham, E.~J.~Yoo and S.~K.~Oh,
  Phys.\ Rev.\ D {\bf 76}, 075011 (2007);~
%
  S.~W.~Ham and S.~K.~Oh,
  Phys.\ Rev.\ D {\bf 76}, 095018 (2007);~
%
  A.~Ahriche and S.~Nasri,
  Phys.\ Rev.\ D {\bf 83}, 045032 (2011).

\bibitem{EWBG_sMSSM}
  J.~Kang, P.~Langacker, T.~-j.~Li and T.~Liu,
  Phys.\ Rev.\ Lett.\  {\bf 94}, 061801 (2005);~
%
  J.~Kang, P.~Langacker, T.~Li and T.~Liu,
  JHEP {\bf 1104}, 097 (2011)
  [arXiv:0911.2939 [hep-ph]].

\bibitem{Chiang:2009fs} 
  C.~-W.~Chiang and E.~Senaha,
  JHEP {\bf 1006}, 030 (2010)
  [arXiv:0912.5069 [hep-ph]].
%

\bibitem{Cheung:2012pg} 
  K.~Cheung, T.~-J.~Hou, J.~S.~Lee and E.~Senaha,
  Phys.\ Lett.\ B {\bf 710}, 188 (2012).
  
 \bibitem{126GeVHiggs} 
  G.~Aad {\it et al.}  [ATLAS Collaboration],
  Phys.\ Lett.\ B {\bf 716}, 1 (2012);~
%
  S.~Chatrchyan {\it et al.}  [CMS Collaboration],
  Phys.\ Lett.\ B {\bf 716}, 30 (2012).

\bibitem{Patel:2011th} 
  H.~H.~Patel and M.~J.~Ramsey-Musolf,
  JHEP {\bf 1107}, 029 (2011).

\bibitem{Wainwright:2012zn} 
  C.~L.~Wainwright, S.~Profumo and M.~J.~Ramsey-Musolf,
  Phys.\ Rev.\ D {\bf 86}, 083537 (2012).

\bibitem{Garny:2012cg} 
  M.~Garny and T.~Konstandin,
  JHEP {\bf 1207}, 189 (2012).
  
\bibitem{Langacker:2008yv} 
  P.~Langacker,
  Rev.\ Mod.\ Phys.\  {\bf 81}, 1199 (2009).

\bibitem{Miller:2003ay} 
  D.~J.~Miller, 2, R.~Nevzorov and P.~M.~Zerwas,
  Nucl.\ Phys.\ B {\bf 681}, 3 (2004).

\bibitem{Coleman:1973jx} 
  S.~R.~Coleman and E.~J.~Weinberg,
  Phys.\ Rev.\ D {\bf 7}, 1888 (1973).

\bibitem{Funakubo:2009eg}
  K.~Funakubo and E.~Senaha,
  Phys.\ Rev.\ D {\bf 79} (2009) 115024
  [arXiv:0905.2022 [hep-ph]].

\bibitem{Arnold:1987mh} 
  P.~B.~Arnold and L.~D.~McLerran,
  Phys.\ Rev.\ D {\bf 36}, 581 (1987).

\bibitem{sph_SM}
  N.~S.~Manton,
  Phys.\ Rev.\ D {\bf 28}, 2019 (1983);~
%
  F.~R.~Klinkhamer and N.~S.~Manton,
  Phys.\ Rev.\ D {\bf 30}, 2212 (1984).

\bibitem{Senaha:2013fva} 
  E.~Senaha,
  arXiv:1305.1563 [hep-ph].

\bibitem{Funakubo:2005bu} 
  K.~Funakubo, A.~Kakuto, S.~Tao and F.~Toyoda,
  Prog.\ Theor.\ Phys.\  {\bf 114}, 1069 (2006)
  [hep-ph/0506156].

\bibitem{DeSimone:2011ek} 
  A.~De Simone, G.~Nardini, M.~Quiros and A.~Riotto,
  JCAP {\bf 1110}, 030 (2011)
  [arXiv:1107.4317 [hep-ph]].

\bibitem{CTP_Riotto}
  A.~Riotto,
  Nucl.\ Phys.\ B {\bf 518}, 339 (1998);~
%
  A.~Riotto,
  Phys.\ Rev.\ D {\bf 58}, 095009 (1998).
  
\bibitem{Lee:2004we} 
  C.~Lee, V.~Cirigliano and M.~J.~Ramsey-Musolf,
  Phys.\ Rev.\ D {\bf 71}, 075010 (2005)
  [hep-ph/0412354].

\bibitem{Cohen:1994ss} 
  A.~G.~Cohen, D.~B.~Kaplan and A.~E.~Nelson,
  Phys.\ Lett.\ B {\bf 336}, 41 (1994)
  [hep-ph/9406345].
 
\bibitem{Huet:1995sh} 
  P.~Huet and A.~E.~Nelson,
  Phys.\ Rev.\ D {\bf 53}, 4578 (1996)
  [hep-ph/9506477].
 
\bibitem{Cline:2000nw} 
  J.~M.~Cline, M.~Joyce and K.~Kainulainen,
  JHEP {\bf 0007}, 018 (2000)
  [hep-ph/0006119].
 
\bibitem{Giudice:1993bb} 
  G.~F.~Giudice and M.~E.~Shaposhnikov,
  Phys.\ Lett.\ B {\bf 326}, 118 (1994)
  [hep-ph/9311367].
 
 
%

\bibitem{Ade:2013zuv} 
  P.~A.~R.~Ade {\it et al.}  [Planck Collaboration],
  arXiv:1303.5076 [astro-ph.CO].

\bibitem{Moore:1995si} 
  G.~D.~Moore and T.~Prokopec,
  Phys.\ Rev.\ D {\bf 52}, 7182 (1995)
  [hep-ph/9506475].

\bibitem{vw_LSS}
  P.~John and M.~G.~Schmidt,
  Nucl.\ Phys.\ B {\bf 598}, 291 (2001)
  [Erratum-ibid.\ B {\bf 648}, 449 (2003)];~
%
  S.~J.~Huber and M.~Sopena,
  Phys.\ Rev.\ D {\bf 85}, 103507 (2012)
  [arXiv:1112.1888 [hep-ph]].

\bibitem{Moreno:1998bq} 
  J.~M.~Moreno, M.~Quiros and M.~Seco,
  Nucl.\ Phys.\ B {\bf 526}, 489 (1998).

\bibitem{Dobrescu:2013cmh} 
  B.~A.~Dobrescu and F.~Yu,
  arXiv:1306.2629 [hep-ph].

\bibitem{Alitti:1993pn} 
  J.~Alitti {\it et al.}  [UA2 Collaboration],
  Nucl.\ Phys.\ B {\bf 400}, 3 (1993).

\bibitem{Buckley:2011vc} 
  M.~R.~Buckley, D.~Hooper, J.~Kopp and E.~T.~Neil,
  Phys.\ Rev.\ D {\bf 83}, 115013 (2011)
  [arXiv:1103.6035 [hep-ph]].

\bibitem{Umeda:1998nq} 
  Y.~Umeda, G.~-C.~Cho and K.~Hagiwara,
  Phys.\ Rev.\ D {\bf 58}, 115008 (1998)
  [hep-ph/9805447].

\bibitem{Li:2008ez} 
  Y.~Li, S.~Profumo and M.~Ramsey-Musolf,
  Phys.\ Lett.\ B {\bf 673}, 95 (2009)
  [arXiv:0811.1987 [hep-ph]].

\bibitem{gHVV_LHCdata}  
  G.~Aad {\it et al.}  [ATLAS Collaboration],
  arXiv:1307.1427 [hep-ex];~
%
  S.~Chatrchyan {\it et al.}  [CMS Collaboration],
  JHEP {\bf 06}, 081 (2013)
  [arXiv:1303.4571 [hep-ex]].

 
\bibitem{future_LowExp} 
  D.~Boer, M.~Diehl, R.~Milner, R.~Venugopalan, W.~Vogelsang, D.~Kaplan, H.~Montgomery and S.~Vigdor {\it et al.},
  arXiv:1108.1713 [nucl-th];~
%
  J.~Dudek, R.~Ent, R.~Essig, K.~S.~Kumar, C.~Meyer, R.~D.~McKeown, Z.~E.~Meziani and G.~A.~Miller {\it et al.},
  Eur.\ Phys.\ J.\ A {\bf 48}, 187 (2012).

 \bibitem{FCNC_Zp}
  M.~R.~Buckley and M.~J.~Ramsey-Musolf,
  Phys.\ Lett.\ B {\bf 712}, 261 (2012);~
%
  M.~Gonzalez-Alonso and M.~J.~Ramsey-Musolf,
  Phys.\ Rev.\ D {\bf 87}, 055013 (2013).

\bibitem{HL-LHC}
  [ ATLAS Collaboration],
  ``Physics at a High-Luminosity LHC with ATLAS,''
  arXiv:1307.7292 [hep-ex].

\bibitem{ILC}
%
  H.~Baer, T.~Barklow, K.~Fujii, Y.~Gao, A.~Hoang, S.~Kanemura, J.~List and H.~E.~Logan {\it et al.},
  arXiv:1306.6352 [hep-ph].
  
\bibitem{Carrington:1991hz} 
  M.~E.~Carrington,
  Phys.\ Rev.\ D {\bf 45}, 2933 (1992).
\end{thebibliography}
\end{document}